\documentclass[review]{elsarticle}

\usepackage{lineno,hyperref}
\modulolinenumbers[5]

\journal{Journal of Computational Physics}

\usepackage{amsmath,amssymb}
\usepackage{xcolor}
\usepackage{datetime}
\usepackage{subcaption}
\usepackage{setspace}
\usepackage[ruled,vlined,english]{algorithm2e}
\usepackage[babel=true]{csquotes}

\newcommand*\diff{\mathop{}\!\mathrm{d}}
\usepackage{mathrsfs}
\usepackage{amsbsy}
\begin{document}

\begin{frontmatter}

\title{A multigrid/ensemble Kalman filter strategy for assimilation of unsteady flows}

\author[Poitiers]{G. Moldovan\corref{correspondingauthor}}
\cortext[correspondingauthor]{Corresponding author, \textit{gabriel-ionut.moldovan@ensma.fr}}
\author[Poitiers]{G. Lehnasch}
\author[Poitiers]{L. Cordier}
\author[Poitiers]{M. Meldi}
\address[Poitiers]{Institut Pprime, CNRS -
ISAE-ENSMA - Universit\'{e} de Poitiers, 11 Bd. Marie et Pierre Curie,
Site du Futuroscope, TSA 41123, 86073 Poitiers Cedex 9, France}

\begin{abstract}

A sequential estimator based on the Ensemble Kalman Filter for Data Assimilation of fluid flows is presented in this research work. The main feature of this estimator is that the Kalman filter update, which relies on the determination of the Kalman gain, is performed exploiting the algorithmic features of the numerical solver employed as a model. More precisely, the multilevel resolution associated with the multigrid iterative approach for time advancement is used to generate several low-resolution numerical simulations. These results are used as ensemble members to determine the correction via Kalman filter, which is then projected on the high-resolution grid to correct a single simulation which corresponds to the numerical model. The assessment of the method is performed via the analysis of one-dimensional and two-dimensional test cases, using different dynamic equations. The results show an efficient trade-off in terms of accuracy and computational costs required. In addition, a physical regularization of the flow, which is not granted by classical KF approaches, is naturally obtained owing to the multigrid iterative calculations. The algorithm is also well suited for the analysis of unsteady phenomena and, in particular, for potential application to in-streaming Data Assimilation techniques.  

\end{abstract}


\end{frontmatter}



%
%
%
%
%
%
%
%
%

\section{Introduction}
\label{sec:introduction}

In Computational Fluid Dynamics (CFD), newly developed numerical methods are generally assessed in terms of accuracy via comparison with experimental data \cite{Oberkampf_Trucano_PAS_2002}.
In practice, this validation step is far from being trivial since many sources of error are inevitably introduced in the simulations.
First, the partial differential equations used to derive the numerical scheme may be restricted to oversimplified physical models, such as the Boussinesq approximation applied in thermal convection or the incompressibility condition. Second, the discretization process and the use of iterative numerical methods introduce computational errors in the representation of the flow features \cite{Ferziger2002_springer}. Third, boundary and initial conditions are usually very sophisticated in complex applications but detailed \textit{a priori} knowledge is insufficiently available. Last, for very high Reynolds number configurations, turbulence/subgrid-scale modelling must be included in order to reduce the required computational costs  \cite{Pope2000_cambridge}. All of these sources of error exhibit complex interactions owing to the non-linear nature of the dynamical models used in numerical application, such as the Navier-Stokes equations.

The experimental results are also affected by uncertainties and biases. In many cases, the set-up of the problem can be controlled up to a finite precision (alignment between the flow and the wind tunnel or immersed bodies, mass flow rate, \dots). This kind of uncertainty, which is clearly affecting every physical system but cannot be exactly quantified, is usually referred to as \emph{epistemic uncertainty}. In addition, experimental results are also affected by the precision (or bias) of the acquisition and measurement system. Thus, the main difficulty in the comparison between numerical and experimental results is understanding how much of the differences observed is due to an actual lack of precision of the numerical model, and how much is instead associated to differences in the set-up of investigation. 

One possible strategy to obtain an optimal combination of the knowledge coming from  simulations and experiments is to derive a state estimation which complies with both sources of information. The degree of precision of such estimation is connected with the confidence in the sources of information.
This  has the advantage of naturally incorporating uncertainty and bias present in the sources of information in the analysis. Tools providing such state estimation are usually included in different disciplines, control theory for state observers \cite{Zhou_Doyle_Glover_1996}, 
Data Assimilation (DA) \cite{Asch2016_SIAM,Evensen2009_Springer,Daley1991_cambridge} for weather prediction, ocean modelling and more recently  mechanical engineering problems. Essentially, DA methods combine information issued from two sources: i) a \emph{model}, which provides a dynamical description of the phenomenon in the physical domain, ii) a set of \emph{observations}, which are usually sparse and/or local in time and space.  These methods are classified  in different families according to the way the state estimation is performed. One of the classical criterion of classification deals with the operative strategy used to derive the state estimation. \emph{Variational} approaches resolve a constrained optimization problem over the parametric space characterizing the model (usually coefficients defining boundary conditions or physical models). The solution of the variational problem minimizes prescribed error norms so that the assimilated model complies with the observation provided over a specified time window. Methods from this family, which include $3$D-Var and $4$D-Var, usually exhibit very high accuracy \cite{Onder2016_cf,Foures2014_jfm,Mons2019_jcp,Chandramouli_Memin_Heitz_JCP_2020}. However, they are also affected by several drawbacks. First, the formulation of the adjoint problem that is introduced to perform the parametric optimization can be difficult, if not impossible, when automatic differentiation is not employed. Second, the adjoint problem is defined backward in time which may lead to numerical difficulties of resolution related to  the unaffordable data storage that is needed, and the amplification of the adjoint solution that frequently happens when multi-scale interactions are dominant \cite{Asch2016_SIAM,Onder2016_cf,Sirkes1997_mwr}. Third, standard variational data assimilation methods are \emph{intrusive} from a computational point of view, requiring the development of an adjoint calculation code, or the availability of the source code  when the use of automatic differentiation is planned.  In most cases, modifications are required in order to maintain the code or extend the applications (change of the definitions of errors, for instance). For commercial software without open-source licence, these modifications are expensive. \emph{Non-intrusive} methods that require no modification of original calculation code are therefore preferable in data assimilation. 

Another family of DA methods is represented by the \emph{sequential} approaches. These methods, mostly based on Bayes' theorem, provide a probabilistic description of the state estimation. A well known approach is the Kalman Filter (KF) \cite{Kalman1960_jbe}. Extensions and generalizations of this method have also been developed, such that the Extended Kalman Filter (EKF) \cite{Welch_Bishop_Report_2006} which is tailored for nonlinear systems, and Ensemble Kalman Filter (EnKF) \cite{Evensen2009_Springer}. This class of methods solves the state estimation problem by transporting error covariance matrices of the model and observations. These methods are usually more flexible than variational approaches (no required computation of first order sensitivities), but the time advancement and update of the covariance matrices are prohibitively expensive for large scale problems, encountered in practical applications \cite{Rozier_Birol_Cosme_Brasseur_Brankart_Verron_SIAM_Review_2007}. One possible strategy consists in reducing the order of the Kalman filter \cite{Suzuki2012_jfm} or filtering the error covariance matrix. Inspired by a domain localization procedure, Meldi \& Poux \cite{Meldi2017_jcp,Meldi2018_ftc} proposed  a strategy based on an explicit filter of the error covariance matrix. The application of this estimator to different turbulent flows exhibited encouraging results considering the relatively small increase in computational resources. A more popular strategy for data assimilation of engineering applications is the Ensemble Kalman Filter \cite{Asch2016_SIAM,Evensen2009_Springer,Evensen2009_IEEE}, which relies on a Monte Carlo implementation of the Bayesian update problem. The EnKF (and follow-up models) was introduced as an extension of the original Kalman filter made for high-dimensional systems for which transporting the covariance matrix is not computationally feasible. EnKF replaces the covariance matrix by the sample covariance matrix computed from an ensemble of state vectors. The main advantage of EnKF is that advancing a high-dimensional covariance matrix is achieved by simply advancing  each member of the ensemble.  Several research works have been reported in the literature in recent years for application in fluid mechanics \cite{Mons2016_jcp,Xiao2016_jcp,Rochoux2014_nhess}. Despite the interest of this non-intrusive technique, and the possibility to perform efficient parametric inference, the computational costs can still be prohibitive for realistic applications. Statistical convergence is usually obtained for a typical ensemble size going from $60$ to $100$ ensemble members \cite{Asch2016_SIAM}.         

In addition, the state estimation obtained via sequential tools does not necessarily comply with a model solution i.e. the \textit{conservativity} of the dynamic equations of the model is violated. This aspect is a potentially critical issue in fluid mechanics studies. Violation of conservativity may result in loss of conservation of some physical properties of the flow (such as mass conservation or momentum conservation) as well as in the emergence of non-physical discontinuities in the flow quantities. The aforementioned issues significantly affect the precision of the prediction of the flow and may eventually produce irreversible instabilities in the time advancement of the dynamical model. A number of works in the literature have provided advancement in the form of additional constraints to be included in the state estimation process. Meldi \& Poux \cite{Meldi2017_jcp} used a recursive procedure and a Lagrangian multiplier (the pressure field) to impose the zero-divergence condition of the velocity field for incompressible flows. Other proposals deal with imposing hard constraints in the framework of an optimization problem \cite{Simon2002_IEEE}, ad-hoc augmented observation \cite{Nachi2007_IEEE} and generalized regularization \cite{Zhang2020_jcp}. These approaches are responsible for a significant increase in the computational resources required, which is due to augmentation in size of the state estimation problem or to the optimization process, which usually needs the calculation of gradients of a cost function.        

The investigation of physically constrained sequential state estimation is here performed using an advanced estimator strategy, which combines an EnKF approach and a multigrid method. 
For this reason, we refer to our algorithm as Multigrid Ensemble Kalman Filter (MEnKF).
Multigrid methods \cite{Brandt1977_mc,Ferziger2002_springer} are a family of tools which employ multi-level techniques to obtain the time-advancement of the flow. In particular, the geometric multigrid \cite{Wesseling1999_jcam} uses different levels of the resolution in the computational grid to obtain the final state. The method here proposed exploits algorithmic features of iterative solvers used in practical CFD applications. The EnKF error covariance matrix reconstruction is performed using information from a number of ensemble members which are generated over a coarse level mesh of a multigrid approach. This procedure is reminiscent of reduced order / multilevel applications of EnKF strategy reported in the literature \cite{Hoel2016_SIAM,Siripatana2019_cg,Fossum2020_cg,Brajard2020_arxiv}. However, the state estimation obtained at the coarse level is used to obtain a single solution calculated on a high resolution mesh grid, similarly to the work by Debreu et al. \cite{Debreu2015_qjrms} for variational DA. Because of the algorithmic structure of the problem, all of the simulations on the fine and coarse level can be run simultaneously in parallel calculations, providing a tool able to perform in-streaming DA for unsteady flow problems.
 
The strategy is tested on different configurations, using several physical models represented by the Burger's equation and the compressible Euler and Navier-Stokes equations for one-dimensional and two-dimensional test cases.

The article is structured as follows. In Sec.~\ref{sec:maths}, the sequential DA procedure is detailed including descriptions of the classical KF and EnKF methods. The numerical discretization and the multigrid strategy are also presented. In Sec~\ref{sec:multigrid-EnKF}, the algorithm called Multigrid Ensemble Kalman Filter (MEnKF) is discussed. In Sec.~\ref{sec:Burgers1D}, the MEnKF is used to investigate a one-dimensional case using the Burgers' equation. In Sec.~\ref{sec:euler1D}, a second one-dimensional case is investigated, but in this case the dynamical model is represented by a Euler equation. In Sec.~\ref{sec:mixLay2D}, we investigate the case of the two-dimensional compressible Navier-Stokes equations, namely the spatially evolving mixing layer. Finally, in Sec.~\ref{sec:conclusions} concluding remarks are drawn. 

\section{Sequential data assimilation in fluid dynamics}
\label{sec:maths}

In Sec.~\ref{sec:kalmanfilter}, we introduce sequential data assimilation methods starting with the Kalman filter.
The objective is to gradually arrive at the Dual ensemble Kalman Filter methodology (Dual EnKF), which is an essential ingredient of the estimator proposed in this work. 
In Sec.~\ref{sec:discrEq}, the steps that are necessary to transform a general transport equation into a discretized model usable in sequential DA are described. 
A brief description of the multigrid approach employed is also provided. 

\subsection{Sequential data assimilation}
\label{sec:kalmanfilter}
\subsubsection{Kalman filter}
\label{sec:KF}
The role of the Kalman filter (KF) is to provide an estimate of the state of a physical system at time $k$ ($\mathbf{x}_k$), given the initial estimate $\mathbf{x}_0$, a set of measurements or observations, and the information of a dynamical model (e.g., first principle equations):
\begin{equation}
\mathbf{x}_{k}=\mathbf{\mathcal{M}}_{k:k-1} \left(\mathbf{x}_{k-1}, \mathbf{\theta}_k\right)+\mathbf{\eta}_k
\label{eq:x_mdl}
\end{equation} 
where $\mathbf{\mathcal{M}}_{k:k-1}$ is a non linear function acting as state-transition model and $\mathbf{\theta}_k$ contains the parameters that affect the state-transition. The term $\mathbf{\eta}_k$ is associated with uncertainties in the model prediction which, as discussed before, could emerge for example from incomplete knowledge of initial / boundary conditions. In the framework of KF applications, these uncertainties are usually modelled as a zero-mean Gaussian distribution characterized by a variance $\mathbf{Q}_k$, \textit{i.e.} $\mathbf{\eta}_k\sim\mathcal{N}(\mathbf{0},\mathbf{Q}_k)$.

Indirect observations of $\mathbf{x}_k$ are available in the components of the observation vector $\mathbf{y}_k^\text{o}$. These two variables are related by:
\begin{equation}
\mathbf{y}_k^\text{o}=\mathbf{\mathcal{H}}_k (\mathbf{x}_k) + \mathbf{\epsilon}_k^\text{o}
\label{eq:z_mdl}
\end{equation}
where $\mathbf{\mathcal{H}}_k$ is the non linear observation operator which maps the model state space to the observed space. The available measurements are also affected by uncertainties which are assumed to follow a zero-mean Gaussian distribution characterized by a variance $\mathbf{R}_k$, \textit{i.e.} $\mathbf{\epsilon}_k^\text{o}\sim\mathcal{N}(\mathbf{0},\mathbf{R}_k)$.

The model and observation errors being Gaussian, we can show that the solution is characterized entirely by the first two moments of the state.
Following the notation generally used in DA literature, the \textit{forecast}/\textit{analysis} states and error covariances are indicated as $\mathbf{x}^{\text{f}/\text{a}}_k$ and $\mathbf{P}^{\text{f}/\text{a}}_k$, respectively.
The error covariance matrix is defined as 
$\mathbf{P}^{\text{f}/\text{a}}_k=
\mathbb{E}\left[
\left(\mathbf{x}_k^{\text{f}/\text{a}} - \mathbb{E}(\mathbf{x}_k^{\text{f}/\text{a}})\right) 
\left(\mathbf{x}_k^{\text{f}/\text{a}} - \mathbb{E}(\mathbf{x}_k^{\text{f}/\text{a}})\right)^\top \right]$.
In the case of a linear dynamical model ($\mathbf{\mathcal{M}}_{k:k-1}\equiv\mathbf{M}_{k:k-1}$) and a linear observation model ($\mathbf{\mathcal{H}}_k\equiv\mathbf{H}_k$), the estimated state is obtained via the following recursive procedure:
\begin{enumerate}
  \item A predictor (forecast) phase, where the analysed state of the system at a previous time-step is used  to obtain an \textit{a priori} estimation of the state at the current instant. This prediction, which is obtained relying on the model only, is not conditioned by observation at time $k$:
\begin{eqnarray}
\mathbf{x}^\text{f}_{k}&=&\mathbf{M}_{k:k-1} \mathbf{x}^\text{a}_{k-1}\label{eq:x_mdl_kf}\\
\mathbf{P}^\text{f}_{k}&=&\mathbf{M}_{k:k-1} \mathbf{P}^\text{a}_{k-1} {\mathbf{M}^\top_{k:k-1}}+\mathbf{Q}_k \label{eq:P_mdl_kf}
\end{eqnarray}
  \item An update (analysis) step, where the state estimation is updated accounting for observation at the time $k$:
   \begin{eqnarray}
  \mathbf{K}_k&=&\mathbf{P}^\text{f}_{k}{\mathbf{H}^\top_k}\left(\mathbf{H}_k \mathbf{P}^\text{f}_{k} {\mathbf{H}^\top_k}+\mathbf{R}_k\right)^{-1}\label{eq:k_kf}\\
  \mathbf{x}^\text{a}_{k}&=&\mathbf{x}^\text{f}_{k} + \mathbf{K}_k\left(\mathbf{y}^\text{o}_k-\mathbf{H}_k\mathbf{x}^\text{f}_{k}\right)\label{eq:x_kf}\\
  \mathbf{P}^\text{a}_{k}&=&\left(I-\mathbf{K}_k \mathbf{H}_k\right)\mathbf{P}^\text{f}_{k}\label{eq:P_kf}
  \end{eqnarray}
\end{enumerate}
The optimal prediction of the state ($\mathbf{x}^\text{a}_{k}$) is obtained via the addition to the predictor estimation ($\mathbf{x}^\text{f}_{k}$) of a correction term determined via the so called \textit{Kalman gain} $\mathbf{K}_k$. 
The classical KF algorithm is not suited for direct application to the analysis of complex flows. First of all, KF classical formulation is developed for linear systems. Applications to non-linear systems can be performed using more advanced techniques such as the extended Kalman filter \cite{Welch_Bishop_Report_2006} or exploiting features of the numerical algorithms used for numerical discretization \cite{Meldi2017_jcp}.
The canonical Kalman filter is difficult to implement with realistic engineering models because of i) nonlinearities in the models and observation operators, ii) poorly known error statistics and iii) a prohibitive computational cost.
Considering point ii), the matrices $\mathbf{Q}_k$ and $\mathbf{R}_k$ are usually unknown and their behaviour must be modelled. One simple, classical simplification is to consider that errors for each component are completely uncorrelated in space and from other components \textit{i.e.} $\mathbf{Q}_k$ and $\mathbf{R}_k$ are considered to be diagonal \cite{Stonebridge_MSc_2017,Brunton2015_amr}.
Point iii) is easily explained by the fact that in engineering applications, the dimension of the physical state $\mathbf{x}_k$ is huge ($N \in [10^6, \, 10^9]$). 
By examining the equations of the canonical Kalman filter, it is immediate to see that the procedure relies on the transport of a very large error covariance matrix $\mathbf{P}_k$. It is therefore necessary to store it but also to invert very large matrices (see \eqref{eq:k_kf}). 
To bypass this computational cost, one popular strategy is to rely on a statistical description obtained via an ensemble of 
trajectories of the model dynamics.
\subsubsection{Ensemble Kalman filter}
\label{sec:EnKF}
The Ensemble Kalman Filter (EnKF) \cite{Evensen2009_Springer}, introduced by Evensen in 1994 \cite{Evensen1994}, relies on the estimation  of $\mathbf{P}_k$ by means of an ensemble. More precisely, the error covariance matrix is approximated using a finite ensemble of model states of size $N_\text{e}$. If the ensemble members are generated using stochastic Monte-Carlo sampling, the error in the approximation decreases with a rate of $\displaystyle\frac{1}{\sqrt{N_\text{e}}}$. In the following, we are describing  the \emph{stochastic EnKF} flavour  of EnKF for which random sampling noise is introduced in the analysis \cite{Asch2016_SIAM}. 

Given an ensemble of forecast/analysed states at a certain instant $k$, the ensemble matrix is defined as:
\begin{equation}
\pmb{\mathscr{E}}_k^{\text{f}/\text{a}}=
\left[\mathbf{x}_k^{\text{f}/\text{a},(1)},\cdots,\mathbf{x}_k^{\text{f}/\text{a},(N_\text{e})}\right]\in\mathbb{R}^{N_x\times N_\text{e}} \label{eq:Ensemble_Matrix}
\end{equation}

To reduce the numerical cost of implementation, the normalized ensemble anomaly matrix is then specified as:
\begin{equation}
\mathbf{X}_k^{\text{f}/\text{a}}=\frac{\left[\mathbf{x}_k^{\text{f}/\text{a},(1)}-\overline{\mathbf{x}_k^{\text{f}/\text{a}}},\cdots,\mathbf{x}_k^{\text{f}/\text{a},(N_\text{e})}-\overline{\mathbf{x}_k^{\text{f}/\text{a}}}\right]}{\sqrt{N_\text{e}-1}}\in\mathbb{R}^{N_x\times N_\text{e}}, \label{eq:Ensemble_Anomaly}
\end{equation}
where the ensemble mean $\overline{\mathbf{x}_k^{\text{f}/\text{a}}}$ is obtained as:
\begin{equation}
    \overline{\mathbf{x}_k^{\text{f}/\text{a}}}=\frac{1}{N_\text{e}}\sum^{N_\text{e}}_{i=1}\mathbf{x}_k^{\text{f}/\text{a},(i)} \label{eq:Ensemble_Mean}
\end{equation}

The error covariance matrix  $\mathbf{P}_k^{\text{f}/\text{a}}$ can thus be estimated via the information derived from the ensemble.
This estimation, hereafter denoted with the superscript $e$, can be factorized into: 
\begin{equation}
\mathbf{P}_k^{\text{f}/\text{a},\text{e}}=\mathbf{X}_k^{\text{f}/\text{a}} \left(\mathbf{X}_k^{\text{f}/\text{a}}\right)^\top\in\mathbb{R}^{N_x\times N_x} \label{eq:Ensemble_P}
\end{equation}

The goal of the EnKF is to mimic the BLUE (Best Linear Unbiased Estimator) analysis of the Kalman filter.
For this,  Burgers et al. \cite{Burgers1998} showed that the observation must be considered as a random variable with an average corresponding to the observed value and a covariance $\mathbf{R}_k$ (the so-called \emph{data randomization} trick).
Therefore, given the discrete observation vector $\mathbf{y}^\text{o}_k\in \mathbb{R}^{N_y}$ at an instant $k$, the ensemble of perturbed observations is defined as:
\begin{equation}
\mathbf{y}_k^{\text{o},(i)}=\mathbf{y}_k^\text{o}+\mathbf{\epsilon}_k^{\text{o},(i)},
\quad
\text{with}
\quad
i=1,\cdots,N_\text{e}
\quad
\text{and}
\quad
\mathbf{\epsilon}_k^{\text{o},(i)}\sim \mathcal{N}(0,\mathbf{R}_k)
.\label{eq:perturbed_y}
\end{equation}

In a similar way to what we did for the ensemble matrices of the forecast/analysed states, we define the normalized anomaly matrix of the observations errors as
\begin{equation}
\mathbf{E}_k^{\text{o}}=
\frac{1}{\sqrt{N_\text{e}-1}}
\left[
\mathbf{\epsilon}_k^{\text{o},(1)}-\overline{\mathbf{\epsilon}_{k}^{\text{o}}},
\mathbf{\epsilon}_k^{\text{o},(2)}-\overline{\mathbf{\epsilon}_{k}^{\text{o}}},
\cdots,
\mathbf{\epsilon}_k^{\text{o},(N_\text{e})}-\overline{\mathbf{\epsilon}_{k}^{\text{o}}},
\right]
\in\mathbb{R}^{N_y\times N_\text{e}}
\end{equation} 
where
$\displaystyle\overline{\mathbf{\epsilon}_{k}^{\text{o}}}=\frac{1}{N_\text{e}}\sum_{i=1}^{N_\text{e}}\mathbf{\epsilon}_k^{\text{o},(i)}$.

The covariance matrix of the measurement error can then be estimated as
\begin{equation}
\mathbf{R}_k^\text{e}=\mathbf{E}^\text{o}_k \left(\mathbf{E}^\text{o}_k\right)^\top\in\mathbb{R}^{N_y\times N_y}. \label{eq:observation_matrix_errorcov}
\end{equation}

By combining the previous results, we obtain (see \cite{Asch2016_SIAM}) the standard stochastic EnKF algorithm. The corresponding analysis step consists of updates  performed on each of the ensemble members, as given by 
\begin{equation}
\mathbf{x}_k^{\text{a},(i)}=
\mathbf{x}_k^{\text{f},(i)}+
\mathbf{K}_k^\text{e}
\left(y_k^{\text{o},(i)}-\mathbf{\mathcal{H}}_k\left(\mathbf{x}^{\text{f},(i)}_k\right)\right)\label{eq:ensemble_update}
\end{equation}
The expression of the Kalman gain is
\begin{equation}
\mathbf{K}_k^\text{e}=
\mathbf{X}_k^\text{f}
\left(\mathbf{Y}_k^\text{f}\right)^\top
\left(
\mathbf{Y}_k^\text{f}
\left(\mathbf{Y}_k^\text{f}\right)^\top
+
\mathbf{E}_k^\text{o} \left(\mathbf{E}_k^\text{o}\right)^\top
\right)^{-1}
\end{equation}
where $\mathbf{Y}_k^\text{f}=\mathbf{H}_k\mathbf{X}_k^\text{f}$.

Interestingly, it can be shown that the computation of the tangent linear operator $\mathbf{H}_k$ can be avoided. The details can be found in \cite[][Sec.~6.3.3]{Asch2016_SIAM}.  A version of the Ensemble Kalman filter algorithm using the previously defined anomaly matrices is given in \ref{sec:EnKFalgo}. This is the version we use in our applications.

State-of-the-art approaches based on the EnKF are arguably the most advanced forms of state estimation available in the field of DA methods. These techniques have been extensively applied in the last decade in meteorology and geoscience \cite{Asch2016_SIAM}. Applications in mechanics and engineering are much more recent, despite a rapid increase in the number of applications in the literature. Among those, studies dealing with wildfire propagation \cite{Rochoux2014_nhess}, combustion \cite{Labahn2019_pci}, turbulence modeling \cite{Xiao2016_jcp} and hybrid variational-EnKF methods \cite{Mons2019_jcp} have been reported. These applications reinforce the idea that approaches based on EnKF have a high investigative potential despite the highly non-linear, multiscale features of the flows studied by the fluid mechanics community.

\subsubsection{Dual Ensemble Kalman filter}
\label{sec:DualEnKF}
In this section, we extend the classical EnKF framework presented in Sec.~\ref{sec:EnKF} by considering the case of a parameterized model such as \eqref{eq:x_mdl}. The objective is to enable the model to generate accurate forecasts. For this, we need to determine good estimates of both model state variables $\mathbf{x}_{k}$ and parameters $\mathbf{\theta}_k$ given erroneous observations $\mathbf{y}_k^{\text{o}}$. One approach is provided by \emph{joint estimation} where state and parameter vectors are concatenated into a single joint state vector (state augmentation). After \cite{DENKF_MORADKHANI2005}, the drawback of such strategy is that, by increasing the number of unknown model states and parameters, the degree of freedom in the system increases and makes the estimation unstable and intractable, especially in the non linear dynamical model. For this reason, we follow the procedure developed by \cite{DENKF_MORADKHANI2005}, called \emph{dual estimation}. The principle is to apply successively two interactive filters, one for the estimation of the parameters from a guessed state solution, the other for the updating of the state variables from the estimated previous parameters.

In the first step of the algorithm, the ensemble of the analysed parameters is updated following the classical KF equation:
\begin{equation}
\mathbf{\theta}_{k}^{\text{a},(i)}  =
\mathbf{\theta}_{k}^{\text{f},(i)}+
\mathbf{K}_k^{\theta,\text{e}}
\left(
\mathbf{y}_k^{\text{o},(i)}-\mathbf{y}_{k}^{\text{f},(i)}
\right)
\quad
\text{with}
\quad
i=1,\cdots,N_e
\label{eq:EnKF par correction}
\end{equation}
where $\mathbf{y}_{k}^{\text{f},(i)} = \mathbf{\mathcal{H}}_k\left(\mathbf{x}_{k}^{\text{f},(i)}\right)$.

The Kalman gain responsible for correcting the parameter trajectories in the ensemble is obtained as follows:
\begin{equation}
\mathbf{K}_k^{\theta,\text{e}}  = 
\mathbf{\Theta}_k^\text{f}
\left(\mathbf{Y}_k^\text{f}\right)^\top
\left(
\mathbf{Y}_k^\text{f}
\left(\mathbf{Y}_k^\text{f}\right)^\top
+
\mathbf{E}_k^\text{o} \left(\mathbf{E}_k^\text{o}\right)^\top
\right)^{-1},
\end{equation}
where the variable $\mathbf{\Theta}_k^\text{f}$ plays the same role for the parameters as the variable $\mathbf{X}_k^{\text{f}}$ defined in \eqref{eq:Ensemble_Anomaly} for the states. We then have:
\begin{equation}
\mathbf{\Theta}_k^{\text{f}/\text{a}}=
\frac{
\left[
\mathbf{\theta}_k^{\text{f}/\text{a},(1)}-\overline{\mathbf{\theta}_k^{\text{f}/\text{a}}},
\cdots,
\mathbf{\theta}_k^{\text{f}/\text{a},(N_\text{e})}-\overline{\mathbf{\theta}_k^{\text{f}/\text{a}}}
\right]
}{
\sqrt{N_\text{e}-1}
}
\in\mathbb{R}^{N_\theta\times N_\text{e}} \label{eq:EnsembleA_Anomaly}
\end{equation}
with
\begin{equation}
\overline{\mathbf{\theta}_k^{\text{f}/\text{a}}}=
\frac{1}{N_\text{e}}
\sum_{i=1}^{N_\text{e}}
\mathbf{\theta}_k^{\text{f}/\text{a},(i)} \label{eq:EnsembleA_Mean}
\end{equation}

Knowing a better approximation of the model's parameters, we can update the state by EnKF (see Sec.~\ref{sec:EnKF}). The Dual Ensemble Kalman filter allows to perform a recursive parametric inference / state estimation using the information from the ensemble members. The algorithm that we use is given in~\ref{sec:DualEnKFalgo}.

\subsection{From transport equation to multigrid resolution}
\label{sec:discrEq}
The general expression for a conservation equation in local formulation over a continuous physical domain reads as:
\begin{equation}
\frac{\mathrm{D} \mathbf{x}}{\mathrm{D} t} = \frac{1}{\rho}\mathbf{\nabla} \cdot \overline{\overline{\sigma}} + \mathbf{f} 
\label{eq:genConsEq}
\end{equation} 
where $\mathrm{D} / \mathrm{D}t$ is the total (or material) derivative of the physical quantity of investigation $\mathbf{x}$  and $\rho$ is the flow density. The divergence operator is indicated as $\mathbf{\nabla} \cdot$ while $\overline{\overline{\sigma}}$ is the stress tensor. Finally, $\mathbf{f}$ represents the effects of volume forces. The evolution of the flow is obtained via time advancement of the discretized solution, which is performed in a physical domain where initial and boundary conditions are provided. A general expression of the discretized form of  \eqref{eq:genConsEq} for the time advancement from the step $k-1$ to $k$ is given by:
\begin{equation}
\mathbf{x}_k = \mathbf{\Phi}_k \mathbf{x}_{k-1} + \mathbf{B}_k \mathbf{b}_k
\label{eq:generalDiscretizedForm}
\end{equation}
where $\mathbf{\Phi}_k$ is the \emph{state transition model} which includes the discretized information of  \eqref{eq:genConsEq}. In case of non-linear dynamics described by \eqref{eq:genConsEq}, the state-of-the-art algorithms used for the discretization process are able to preserve the non-linear information in the product $\mathbf{\Phi}_k \mathbf{x}_{k-1}$, up to a discretization error which is usually proportional to the size of the time step. The term $\mathbf{b}_k$ represents the \textit{control vector} reflecting, for instance, the effect of the boundary conditions. $\mathbf{B}_k$ is the \emph{control input model} which is applied to the control vector $\mathbf{b}_k$. Equation \eqref{eq:generalDiscretizedForm} is consistent with a time explicit discretization of \eqref{eq:genConsEq}. It is well known that this class of methods, despite the very high accuracy, may exhibit some unfavorable characteristics for the simulation of complex flows, such as limitations to the time step according to the Courant-Friedrichs-Lewy (CFL) condition \cite{Ferziger2002_springer}. To bypass this limitation, one possible alternative  consists in using implicit schemes for time discretization. In this case, the general structure of the discretized problem is usually cast in the form:   
\begin{equation}
\mathbf{\Psi}_k \mathbf{x}_k = \widetilde{\mathbf{\Psi}}_k \mathbf{x}_{k-1} + \widetilde{\mathbf{B}}_k \mathbf{b}_k = \mathbf{c}_k
\label{eq:generalDiscretizedFormInt}
\end{equation}
where $\mathbf{\Psi}_k$, $\widetilde{\mathbf{\Psi}}_k$  and $\widetilde{\mathbf{B}}_k$ are matrices obtained via the discretization process. Obviously, considering $\mathbf{\Phi}_k=\mathbf{\Psi}_k^{-1}\widetilde{\mathbf{\Psi}}_k$, we retrieve \eqref{eq:generalDiscretizedForm}. However, this manipulation is in practice not performed due to the prohibitive costs associated to large scale matrices inversions at each time step. 
Instead, an iterative procedure is used until the residual $\mathbf{\delta}^n$, determined at the $n$-th iteration, falls below a pre-selected threshold value $\varepsilon$. 
In other words, the procedure is stopped when $\left\Vert\mathbf{\delta}^n\right\Vert=\left\Vert\mathbf{\Psi}_k \mathbf{x}_k^{n} - \mathbf{c}_k\right\Vert<\varepsilon$.
Among the various iterative methods proposed in the literature, \emph{multigrid} approaches are extensively used in CFD applications \cite{Hackbusch1985_springer,Ferziger2002_springer}. 
The solution is found on the computational grid by updating an initial guess via multiple estimations obtained on a hierarchy of discretizations. 
Two well-known families of multigrid approaches exist, namely the \emph{algebraic} multigrid method and the \textit{geometric} multigrid method.
With algebraic multigrid methods, a hierarchy of operators is directly constructed from the state transition model $\mathbf{\Psi}$. 
On the other hand, the geometric multigrid obtains the solution via a set of operations performed in two (or more) meshes. 
In this paper, we consider the simplified case of two grids. Thereafter, the variables defined on the fine grid will be denoted with the superscript F ($\mathbf{x}^\text{\tiny F}$ for instance), those defined on the coarse grid will be denoted with the superscript C ($\mathbf{x}^\text{\tiny C}$ for instance).

The coarse-level representation $\mathbf{x}^\text{\tiny C}$ is usually obtained suppressing multiple mesh elements from the initial fine-level one $\mathbf{x}^\text{\tiny F}$. This operation may be defined by a coarsening ratio parameter $r_\text{\tiny C}$, which indicates the total number of elements on the fine grid over the number of elements conserved in the coarse grid. Among the numerous algorithms proposed for geometric multigrid, we use the Full Approximation Scheme (FAS), which is a well documented strategy \cite{Brandt1977_mc,Wesseling1999_jcam}. A general formulation for a two-grid algorithm is now provided. The time subscript $k$ is excluded for clarity. The superscript $n$ represents the iteration step of the procedure.
 
\begin{enumerate}
\item Starting from an initial solution on the fine grid $\left(\mathbf{x}^0\right)^\text{\tiny F}$ (which is usually equal to $\mathbf{x}$ at the previous time step $k-1$), an iterative procedure is applied to obtain a first solution $\left(\mathbf{x}^1\right)^\text{\tiny F}$. A residual $\left(\mathbf{\delta}^1\right)^\text{\tiny F}= \mathbf{c}^\text{\tiny F} - \mathbf{\Psi}^\text{\tiny F} \left(\mathbf{x}^1\right)^\text{\tiny F}$ is calculated.
\item $\left(\mathbf{x}^1\right)^\text{\tiny F}$ and $\left(\mathbf{\delta}^1\right)^\text{\tiny F}$ are projected from the fine grid to the coarse grid space via a projection operator $\Pi_\text{\tiny C}$, so that $\left(\mathbf{x}^1\right)^\text{\tiny C}$ and $\left(\mathbf{\delta}^1\right)^\text{\tiny C}$ are obtained. Similarly, the state transition model $\mathbf{\Psi}^\text{\tiny F}$ is projected on the coarse grid (that is re-estimated based on the projection of the solution of the fine grid onto the coarse grid) to obtain $\mathbf{\Psi}^\text{\tiny C}$. Finally, we evaluate  $\mathbf{c}^\text{\tiny C}=\mathbf{\Psi}^\text{\tiny C}\,\left(\mathbf{x}^1\right)^\text{\tiny C}+\left(\mathbf{\delta}^1\right)^\text{\tiny C}$.
\item An iterative procedure is employed to obtain $\left(\mathbf{x}^2\right)^\text{\tiny C}$ on the coarse grid using as initial solution $\left(\mathbf{x}^1\right)^\text{\tiny C}$.
\item The updated variable on the fine grid is obtained as $\left(\mathbf{x}^2\right)^\text{\tiny F}=\left(\mathbf{x}^1\right)^\text{\tiny F}+\Pi_\text{\tiny F}\left(\left(\mathbf{x}^2\right)^\text{\tiny C}-\left(\mathbf{x}^1\right)^\text{\tiny C}\right)$ where $\Pi_\text{\tiny F}$ is a projection operator from the coarse grid to the fine grid.
\item At last, the final solution $\left(\mathbf{x}^3\right)^\text{\tiny F}$ is obtained via a second iterative procedure on the fine grid starting from the intermediate solution $\left(\mathbf{x}^2\right)^\text{\tiny F}$. 
\end{enumerate}

This procedure can be repeated multiple times imposing $\left(\mathbf{x}^0\right)^\text{\tiny F}=\left(\mathbf{x}^3\right)^\text{\tiny F}$ at the beginning of each cycle. When the convergence is reached, the fine grid solution at time instant $k$ is equal to 
$\left(\mathbf{x}^3\right)^\text{\tiny F}$. Performing part of the calculations on a coarse grid level provides two main advantages \cite{Ferziger2002_springer}. First, a significant reduction in the computational resources is obtained
since the calculations performed over the coarse grid are usually much less expensive than a full set of iterations over the fine grid.
Second, spurious high-frequency numerical oscillations, due to the discretization error in $\mathbf{x}^\text{\tiny F}$, are naturally filtered via the operators $\Pi_\text{\tiny C}$ and $\Pi_\text{\tiny F}$, which globally improves the quality of the numerical solution. In this paper, the two projections ($\Pi_\text{\tiny F}$ and $\Pi_\text{\tiny C}$) are defined as 4-th order Lagrange interpolators.

\section{Multigrid Ensemble KF method (MEnKF)}
\label{sec:multigrid-EnKF}
Despite the game-changing advantage that EnKF offers for the analysis of large-scale dynamical systems, the use of a sufficiently large ensemble (usually $60$ to $100$ members are required for convergence \cite{Asch2016_SIAM}) may still be too prohibitive for advanced applications. 
In the following, we present a Kalman filter strategy which relies on the use of a coarse mesh for the ensemble Kalman filter step.
These computations on the coarse mesh are jointly run with a single high-refinement simulation, which is updated using the coarse mesh assimilation results.
For this reason, the clock time required for the time advancement of the ensemble members and the memory storage of the physical variables are dramatically reduced. 
The multigrid-ensemble algorithm works through the steps described below. In the following description, the notation $\mathbf{\Psi}$ might hold for both $\mathbf{\Psi}$ and $\widetilde{\mathbf{\Psi}}$ introduced in \eqref{eq:generalDiscretizedFormInt}, depending on the choice of the time integration strategy. 
For its part, the state $\left(\mathbf{x}^\text{\tiny C}_{k}\right)^{*}$ represents the projection on the coarse-grid of 
$\left(\mathbf{x}_k^\text{\tiny F}\right)^\text{f}$, the fine-grid solution.
Finally, $\left(\mathbf{x}^\text{\tiny C}_k\right)^{'}$ denotes the coarse-grid state obtained by an ensemble  Kalman filtering process applied on the coarse grid.
In the FAS multigrid method described in Sec.~\ref{sec:discrEq}, $\left(\mathbf{x}^\text{\tiny C}_{k}\right)^{*}$ and $\left(\mathbf{x}^\text{\tiny C}_k\right)^{'}$ correspond to  $\left(\mathbf{x}^1\right)^\text{\tiny C}$ and $\left(\mathbf{x}^2\right)^\text{\tiny C}$, respectively.  

\begin{enumerate}
\item \textbf{First iteration on the fine grid}. Starting from an initial solution on the fine grid $\left(\mathbf{x}^\text{\tiny F}_{k-1}\right)^\text{a}$,  a forecasted state $\left(\mathbf{x}^\text{\tiny F}_{k}\right)^\text{f}$ is obtained by using $\overline{\mathbf{\theta}_{k}^\text{a}}$ as parameter for the model $\mathbf{\Psi}^\text{\tiny F}$. 
\item \textbf{Projection on the coarse grid}. $\left(\mathbf{x}^\text{\tiny F}_{k}\right)^\text{f}$ is projected on the coarse grid space via a projection operator $\Pi_\text{\tiny C}$, so that $\left(\mathbf{x}^\text{\tiny C}_{k}\right)^{*}$ is obtained. Similarly, the state matrix $\mathbf{\Psi}^\text{\tiny F}$ is projected on the coarse grid to obtain $\mathbf{\Psi}^\text{\tiny C}$. This critical passage will be discussed in detail in the following. 
\item \textbf{Time advancement of the ensemble members used in the Dual EnKF}. 
For each member $i$ of the ensemble, the state matrix $\left(\mathbf{\Psi}^\text{\tiny C}\right)^{(i)}$ used for the advancement in time on the coarse grid is determined\footnote{%
While the advancement model is unique (the Navier-Stokes equations, for example), the discretization process contained in $\mathbf{\Psi}$ is unique for each member of the ensemble. To distinguish them, it is therefore necessary to introduce an exponent $i$ in the notations. 
}.
The ensemble forecast $\left(\mathbf{x}^\text{\tiny C}_{k}\right)^{\text{f},(i)}$ is corrected with the standard Dual EnKF procedure to obtain $\left(\mathbf{x}^\text{\tiny C}_{k}\right)^{\text{a},(i)}$ as well as the parameters $\mathbf{\theta}_{k}^{\text{a},(i)}$.
\item \textbf{Determination of the state variables on the coarse grid}. If observations are not available, the state $\left(\mathbf{x}^\text{\tiny C}_k\right)^{'}$  is obtained using the classical multigrid procedure. On the other hand, if observations are available, the ensemble error covariance matrix $\left(\mathbf{P}_k^\text{\tiny C}\right)^{\text{f},\text{e}}$
 is used to determine the coarse grid solution $\left(\mathbf{x}^\text{\tiny C}_k\right)^{'}$ through a Kalman filter estimation.
\item \textbf{Final iteration on the fine grid}. A first estimation $\left(\mathbf{x}^\text{\tiny F}_k\right)^{'}$ of the fine grid state is determined using the results obtained on the coarse space: $\left(\mathbf{x}^\text{\tiny F}_k\right)^{'}=\left(\mathbf{x}^\text{\tiny F}_{k}\right)^\text{f}+\Pi_\text{\tiny F}\left(\left(\mathbf{x}^\text{\tiny C}_k\right)^{'}-\left(\mathbf{x}^\text{\tiny C}_{k}\right)^{*}\right)$. The state $\left(\mathbf{x}^\text{\tiny F}_k\right)^\text{a}$ is obtained from a final iterative procedure starting from $\left(\mathbf{x}^\text{\tiny F}_k\right)^{'}$. 
\end{enumerate}

\begin{figure}[htbp]
\centering
\includegraphics[width=1\textwidth]{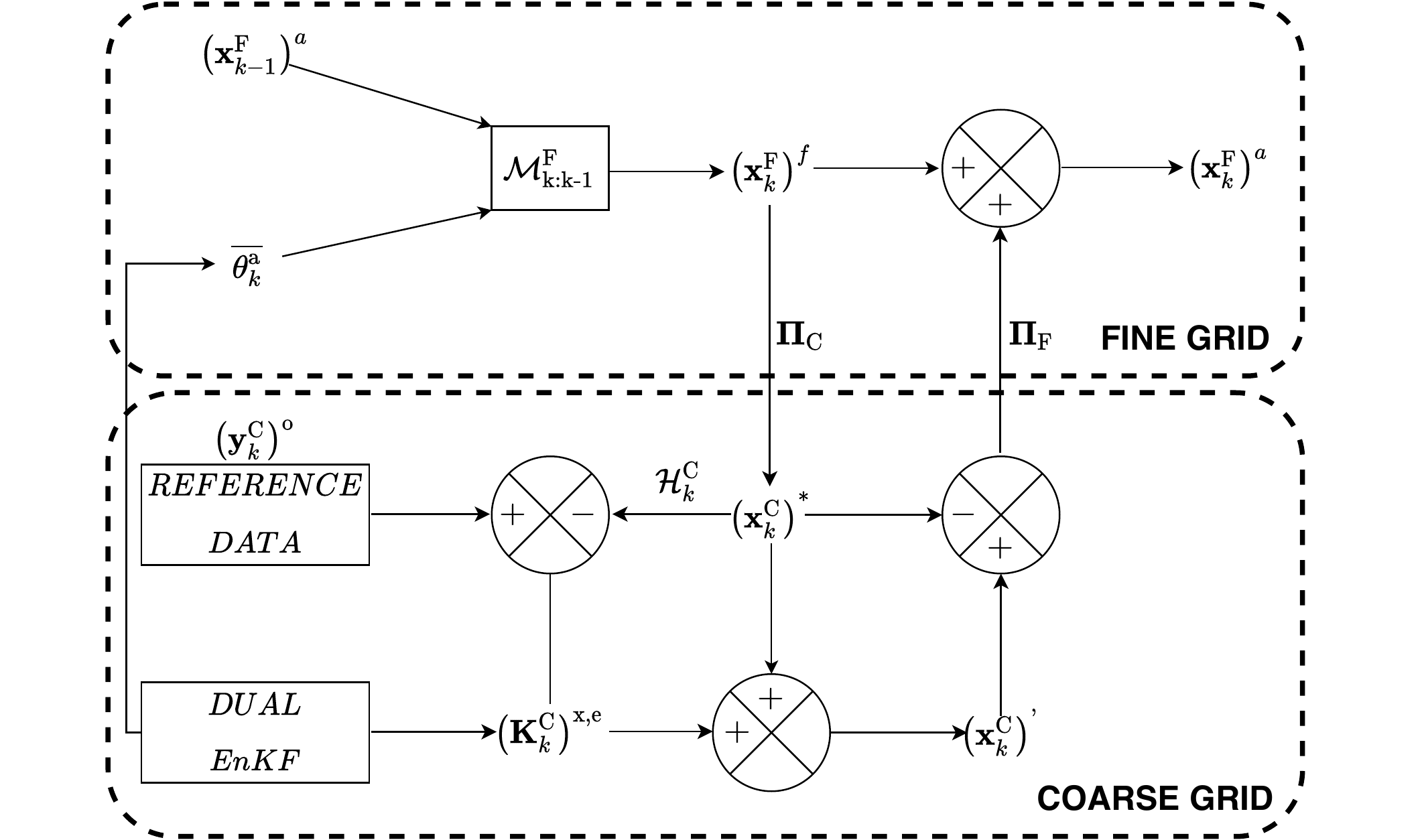}
\caption{\label{fig:overview}
Schematic representation of the Multigrid Ensemble Kalman Filter (MEnKF). 
Two different levels of representation (fine and coarse grids) are used to obtain a data-driven fine grid estimation.
The Dual Ensemble Kalman filter procedure is solved in the coarse grid.
The full algorithm is given in \ref{sec:DA_algorithms}.
}
\end{figure} 

An overview of the assimilation cycle described in the previous algorithm is presented in Fig.~\ref{fig:overview}. Two important aspects need to be discussed:
\begin{itemize}
\item[-] As previously stated, $\mathbf{\Psi}^\text{\tiny C}$ and its role in the determination of the matrices $\left(\mathbf{\Psi}^\text{\tiny C}\right)^{(i)}$ is an essential step in the MEnKF strategy. In non-linear problems of interest in fluid mechanics, the state transition matrix $\mathbf{\Psi}$ includes information of the multi-scale interactions that are specific for every case investigated. The simplest possible choice, which is the one adopted in this work, is to calculate the coefficients of the matrices $\mathbf{\Psi}^\text{\tiny C}$ and $\left(\mathbf{\Psi}^\text{\tiny C}\right)^{(i)}$ separately for each simulated state. Thus, the similarities between the employed state matrices are limited to the use of the same discretization schemes / structure of $\mathbf{\Psi}$. However, one can envision to use the non-linear information conserved in $\mathbf{\Psi}^\text{\tiny C}$, which is supposedly accurate, to improve the accuracy of the prediction of the ensemble members. This aspect is discussed in the perspectives included in Sec.~\ref{sec:conclusions}.
\item[-] The recursive structure of the algorithm allows for integration of iterative corrections for non-linear systems \cite{Sakov2011_mwr} as well as hard constraints (see the discussion in the introduction of \cite{Simon2002_IEEE,Nachi2007_IEEE,Zhang2020_jcp}) to respect the conservativity of the model equations. 
However, these corrections may result in an increase of the computational resources required. Here, the multigrid algorithm itself is used for regularization (\textit{i.e.} for smoothing the discontinuities in the physical variables produced by the update via Kalman Filter) of the flow. In fact, if an intentionally reduced tolerance is imposed in the iterative steps $4$ and $5$, the final solution will keep memory of the features of the state estimation produced in step $3$. However, the iterative resolution will smooth the estimation via the state transition model $\mathbf{\Psi}$, which will perform a natural regularization of the flow. Clearly, if a reduced tolerance is imposed, the final solution will not necessarily respect the conservativity constraints of the model equations. However, one can argue that complete conservativity is not an optimal objective in this case, if the model state at the beginning of the time step is not accurate.      
\end{itemize}

The advantages of our strategy with respect to classical approaches based on EnKF may be summarized in the following points:
\begin{itemize}
\item[-] The RAM requirement necessary to store the $N_\text{e}$ ensemble members during the assimilation is usually moderate. The reduction in computational costs is driven by $N_\text{e}$ and by the size of the coarse variables. 
To illustrate, let us consider the case of a simple two-level geometric multigrid approach for a 3D test case with a constant coarsening ratio $r_\text{\tiny C}=4$ and a size of ensembles $N_\text{e}=100$.
Each ensemble member is then described by $4^3=64$ times less mesh elements than the single simulation on the fine grid. If one considers that one main simulation and $100$ ensemble members are run simultaneously, and if the RAM requirement is normalized over the main simulation, this implies that $R_\text{RAM}$, the non-dimensional RAM requirement, is equal to $1 + 100/64 = 2.56$.
In other words, the total cost in RAM is increased to just $2.56$ times the cost of the simulation without EnKF.
For $r_\text{\tiny C}=8$, the normalized RAM requirement is $R_\text{RAM}=1+100/8^3=1.195$, thus just a $20\%$ increase in RAM requirements. 
This is clearly orders of magnitude more advantageous than a fine-grid classical EnKF application with $N_\text{e}=100$, since in this case $R_\text{RAM}=N_\text{e}=100$.
\item[-] Considering that the ensemble members in the coarse grid and the simulation over the fine grid are running simultaneously, communication times are optimized. 
\item[-] Owing to the iterative procedures of steps $4$ and $5$, regularization of the final solution is naturally obtained. 
\item[-] The algorithm is here described and tested in the framework of geometric multigrid, but it can actually be integrated within other algorithmic structures. For iterative methods, the only essential operation to be performed is the determination of the state transition model $\mathbf{\Psi}^\text{\tiny C}$ and of the projections $\Pi_\text{\tiny C}$ and $\Pi_\text{\tiny F}$. This implies that the method can be easily extended to other popular procedures, such as the algebraic multigrid. If the multigrid operations are removed, the methods becomes a classical multilevel EnKF method, which relies on just one level of resolution for the ensemble members. 
However, in this case, no regularization is obtained unless specific corrections are included. 
\end{itemize}

This general algorithm may be easily tailored accounting for the complexity of the test case investigated, in particular for the requirements of iterative loops on both the coarse grid level and the fine grid level. The algorithm that we used to validate our approach is described in \ref{sec:MEnKF}.

\section{Application: one-dimensional Burgers' equation}
\label{sec:Burgers1D}

\begin{figure}[htbp]
\centering
\includegraphics[width=0.9\textwidth]{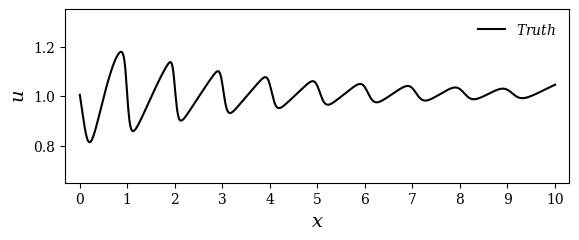}
\caption{\label{fig:1D-Burgers-truth}Solution of the 1D Burgers' equation at $t=10$ for $\theta_1=0.2$ and $\theta_2=0$ (true state).}
\end{figure}

The MEnKF method introduced in Sec.~\ref{sec:multigrid-EnKF} is now applied to the analysis of different test cases. Several dynamical systems of increasing complexity were chosen in order to highlight different properties of the algorithm. Also, a set of different tests is performed in order to obtain a comprehensive validation of the method. At first, let us consider a 1D Burgers' equation:
\begin{equation}
\frac{\partial u}{\partial t}+u\frac{\partial u}{\partial x}=\frac{1}{\text{Re}} \frac{\partial^2 u}{\partial x^2}\label{eq:burgers}
\end{equation}
where $x$ is the spatial coordinate, $u$ the velocity and Re is the Reynolds number. 
Equation \eqref{eq:burgers} is non-dimensionalized with a reference velocity $u_0$ and a reference length $\lambda$.  
This equation is solved with a second-order centered finite difference scheme for the space derivatives and a first-order scheme for the time integration to obtain the general form of discretized representation as given by \eqref{eq:generalDiscretizedFormInt}. A Dirichlet time-varying condition is imposed at the inlet:
\begin{equation}
u(x=0,t)=u_0 \left(1 + \theta_1\sin(\omega t+\theta_2)\right)
\label{eq:inlet-1D-Burgers}
\end{equation}
where $u_0=1$ and $\omega=2\pi f_c$. The characteristic frequency $f_c$ is set to $1$ and the characteristic length $\lambda$  is defined as $\lambda=\frac{u_0}{f_c}$. 
The characteristic time is defined as $t_c=\lambda/u_0$. This implies that, over a complete period of oscillation $1/f_c$, the solution is advected of a length $\lambda$ by the velocity $u_0$. $\theta_\text{1}$ and $\theta_\text{2}$ represent the amplitude and phase of the sinusoidal signal, respectively. Note that from now on, the units are the characteristic magnitudes. The outlet boundary condition is extrapolated from the nearest points to the outlet. The initial condition imposed for $t=0$ is $u(x,t=0)=u_0$ everywhere in the physical domain ($\theta_2=0$ for the reference simulation).
The value of the Reynolds number is $\text{Re}=200$. 
The time advancement step is chosen as $\Delta t = 0.0002$. It is kept constant throughout the simulation. 
The analysis is performed over a physical domain of size $[0, 10]$. 
The distance between the computational nodes in the fine mesh is constant and set to $\Delta x=0.0125$.
This choice has been performed to discretize the characteristic length $\lambda$ using $80$ mesh elements. 
This also implies that the total number of nodes employed to perform the calculation is $N_x=800$.  
A simulation of reference is run on the fine grid with values $\theta_1=0.2$ for the amplitude and $\theta_2=0$ for the phase. 
Thereafter, the solution obtained by this reference simulation is called the \emph{true} state or \emph{truth}.
A flow visualization at $t=10$ is shown in Fig.~\ref{fig:1D-Burgers-truth}. 
For the investigated value of Reynolds number, the non-linear effects and viscous mechanisms can be clearly identified.

Data assimilation is performed in the following conditions:
\begin{itemize}
\item[-] The observations are sampled each $30$ time steps of the reference simulation on the space domain $[0, 1]$ ($80$ sensors) and on the time window $[10, 29]$. The observations are sampled as soon as the flow is fully developed.
These observations are artificially perturbed using a Gaussian noise of variance $R=0.0025$.  
\item[-] The \textit{model} is chosen to be the discretized version of \eqref{eq:burgers}. The numerical test consists of one main simulation, which is run on the fine grid previously introduced, and an ensemble of $N_\text{e}=100$ coarse simulations used for assimilation purposes. The coefficients $\theta_1$ and $\theta_2$ are initially assumed to be described by Gaussian distributions, so that $\theta_1 \sim \mathcal{N}(0, Q_{\theta_1})$ and $\theta_2 \sim \mathcal{N} (0.3, Q_{\theta_2})$. The initial value of the covariance of the parameters is chosen equal to $Q_{\theta_1}(t=0)=Q_{\theta_2}(t=0)=0.0025$. The values prescribed on the fine grid simulation are the mean values of the Gaussian distribution \textit{i.e.} $\theta_1=0$ and $\theta_2=0.3$. Random values for the parameters are imposed at the inlet for each ensemble member on the coarse grid level. The initial mean values for the parameters are significantly different when compared with the values prescribed in the reference simulation, which are $\theta_1=0.2$ and $\theta_2=0$. This choice allows to analyze the rate of convergence of the optimization procedure. Also, the initial condition $u(x,0)=u_0$ is imposed for every simulation.
\end{itemize}
The data estimation is run for a time window of $T_\text{DA}=19$ characteristic times, which encompass roughly $3000$ DA analysis phases.
The sensitivity of the parametric inference procedure to the resolution of the coarse simulations is investigated considering several coarsening ratios $r_\text{\tiny C}=1,\,2,\,4,\,8,\,16$. The fine grid is unchanged, so that the MEnKF is performed using information of progressively coarser grids as $r_\text{\tiny C}$ increases. The interest of this test is to analyse the loss of accuracy of the estimator as $r_\text{\tiny C}$ increases and to ascertain the potential for efficient trade-off between accuracy and computational resources required for the estimation process.

\begin{figure}[htbp]
\includegraphics[width=1\textwidth]{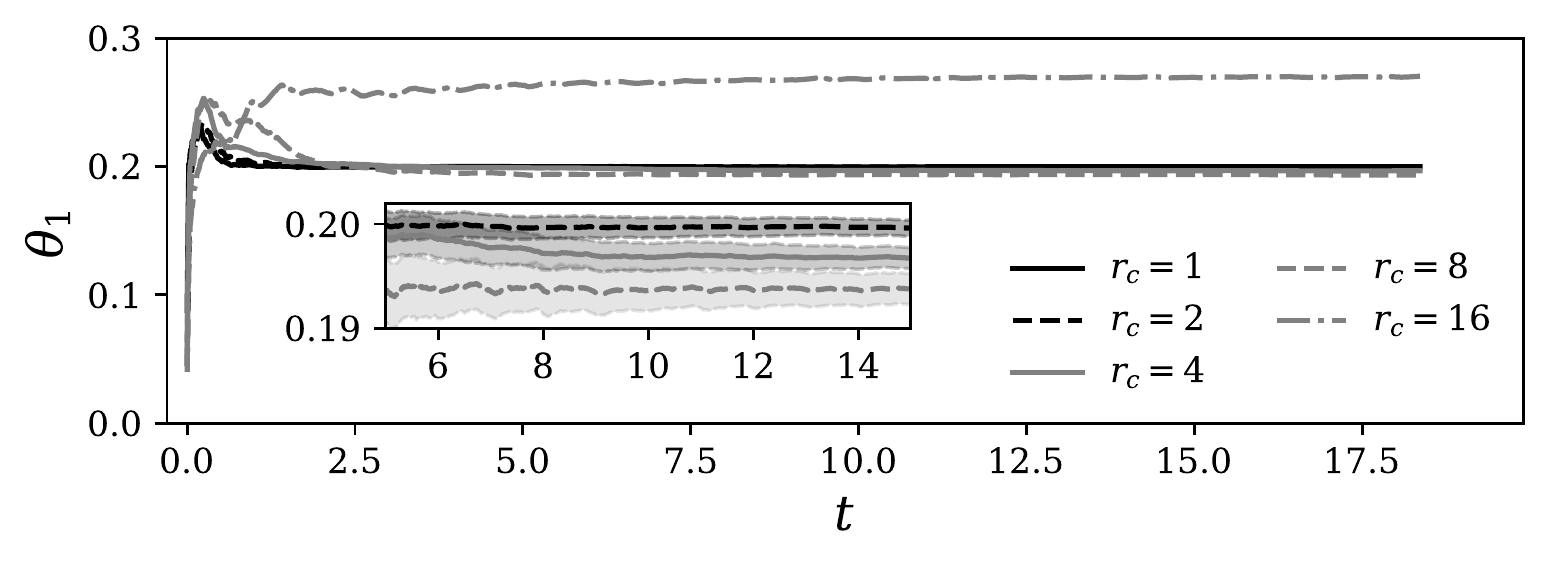}
\caption{\label{fig:amplitude_avg_DENKF}
Values of the parameter $\theta_1$ for different coarsening ratios $r_\text{\tiny C}=1,2,4,8,16$.
In the zoomed region, the shaded area represents the $95\%$ credible interval for the shown cases.}
\end{figure}

The time-evolution of the estimation of $\theta_1$ is shown in Fig.~\ref{fig:amplitude_avg_DENKF}. Very rapid convergence (less than $2 t_c$) is observed for $r_\text{\tiny C} \leq 4$. In addition,  the  parameter  estimation  is  extremely  precise  (discrepancy lower than $0.01\%$ for $r_\text{\tiny C}=1$,  lower than  $2\%$  for $r_\text{\tiny C}=  4$). For  higher values of the parameter $r_\text{\tiny C}$,  the estimation of $\theta_1$ becomes progressively more degraded. For the case $r_\text{\tiny C}=8$, $\theta_1$ is initially overestimated and it finally converges to a value of $\theta_1=0.194$, $3\%$ smaller that the true value. 
Larger errors in the optimization of $\theta_1$ are observed for $r_\text{\tiny C}=16$. In this case, the optimized amplitude parameter is $\theta_1=0.27$, which is $35\%$ larger than the true value. 

\begin{figure}[htbp]
\includegraphics[width=1\textwidth]{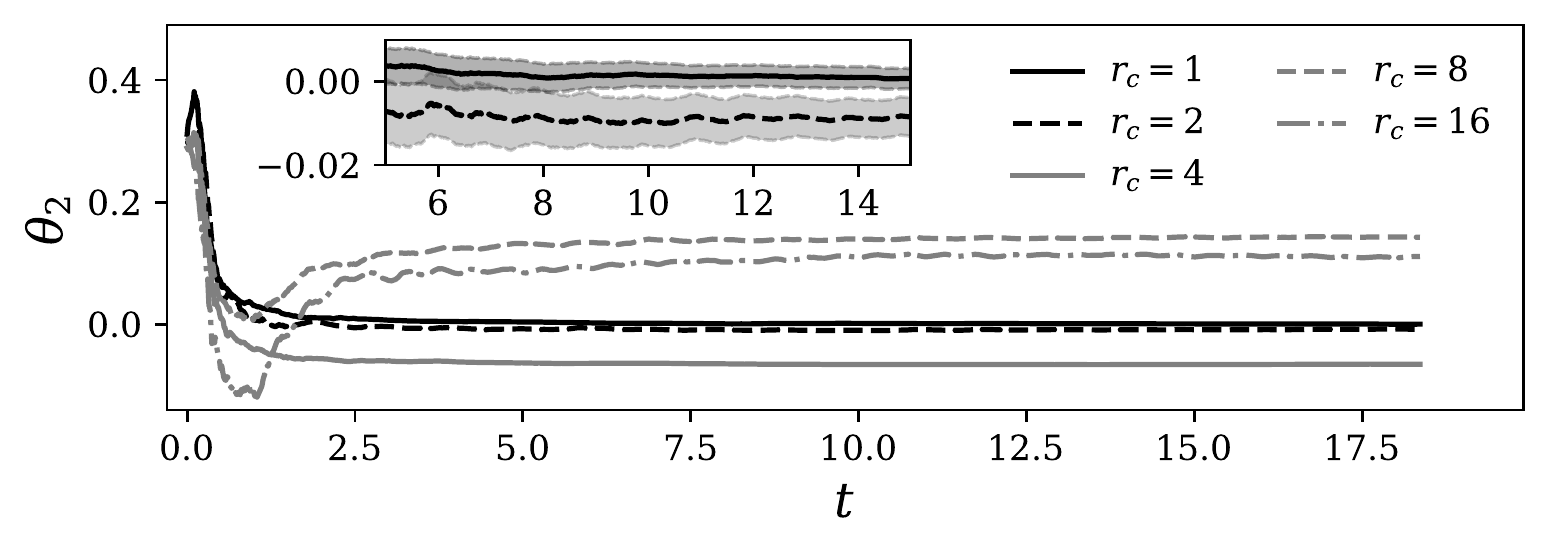}
\caption{\label{fig:phase_avg_DENKF}
Values of the parameter $\theta_2$ for different coarsening ratios $r_\text{\tiny C}=1,2,4,8,16$.
In the zoomed region, the shaded area represents the $95\%$ credible interval for the shown cases.}
\end{figure}

Similar considerations can be drawn by the analysis of the optimization of the parameter $\theta_2$, which is shown in Fig.~ \ref{fig:phase_avg_DENKF}. For $r_\text{\tiny C}=1,\,2,\,4$, we obtain an accurate prediction of the parameter, while a loss in accuracy is observed for the cases $r_\text{\tiny C}=8,16$. This observation can be justified considering the number of mesh elements representing one characteristic length $\lambda$ for these cases, which are $10$ and $5$ for $r_\text{\tiny C}=8$ and $16$, respectively. If one considers that the frequency $\omega$ is set so that one complete oscillatory cycle is performed on average over a characteristic length $\lambda$, this means that the average phase angle between mesh elements is equal to $0.63$ radians for $r_\text{\tiny C}=8$ and $1.26$ radians for $r_\text{\tiny C}=16$. Thus, the values observed for the optimization of $\theta_2$ for these two cases, which are around $\theta_2 \approx 0.15$, are significantly lower than the uncertainty due to the coarse-level resolution.

\begin{figure}[htbp]
  \begin{subfigure}{\textwidth}
  \includegraphics[width=1\linewidth]{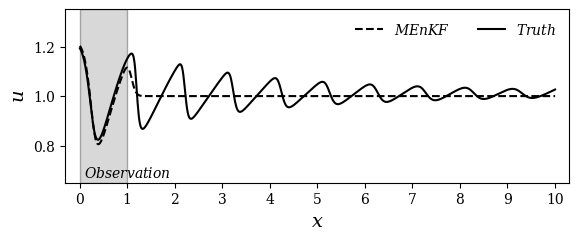}
  \caption{$t=1$}
  \end{subfigure}\par\medskip
  \begin{subfigure}{\textwidth}
  \includegraphics[width=1\linewidth]{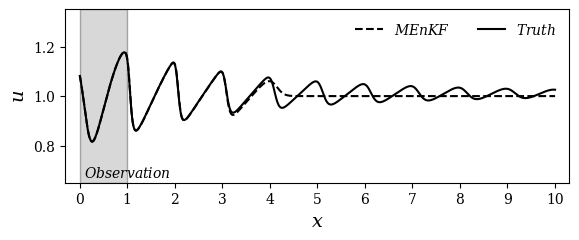}
  \caption{$t=3.88$}
  \end{subfigure}\par\medskip
  \begin{subfigure}{\textwidth}
\includegraphics[width=1\linewidth]{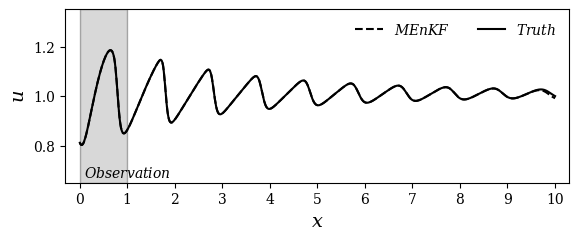}
\caption{$t=10.60$}
  \end{subfigure}
  \caption{\label{fig:MgENKF_states_rc1} 
Estimations obtained by MEnKF for $r_\text{\tiny C}=1$ at
$t=1$ (a), $t=3.88$ (b)  and $t=10.60$ (c). Times are given in $t_c$ units.
The grey shaded area corresponds to the observation window.
}
\end{figure}

State estimation results for the case $r_\text{\tiny C}=1$ are shown in Fig.~\ref{fig:MgENKF_states_rc1}. This case, which is performed using the same grid for the coarse and fine mesh level, is equivalent to a standard Dual EnKF. However, owing to the final multigrid iterative loop, the final solution is naturally regularized. The results, which are shown for $t=1,\,3.88,\,10.60$, show that the estimator successfully represents the behaviour of the dynamical system. A full domain advective time (\textit{i.e.} $10$ characteristic time units) must be simulated in order to observe the effect of the MEnKF in the whole domain. In fact, the parametric information imposed at the inlet for the ensemble members must affect the whole physical domain before a reliable correlation between the state variables can be established. However, once this initial transient is faded, the state estimation almost perfectly captures the behaviour of the true state.  

\begin{figure}[htbp]
  \begin{subfigure}{\textwidth}
  \includegraphics[width=1\linewidth]{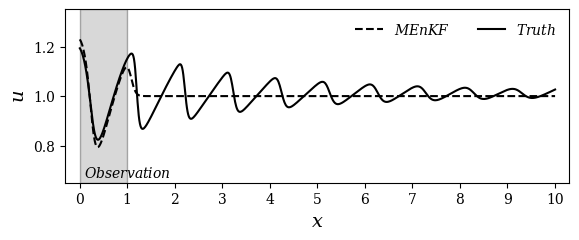}
  \caption{$t=1$}
  \end{subfigure}\par\medskip
  \begin{subfigure}{\textwidth}
  \includegraphics[width=1\linewidth]{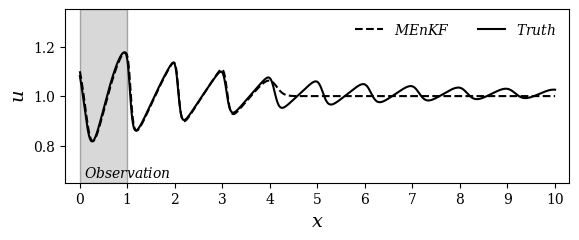}
  \caption{$t=3.88$}
  \end{subfigure}\par\medskip
  \begin{subfigure}{\textwidth}
\includegraphics[width=1\linewidth]{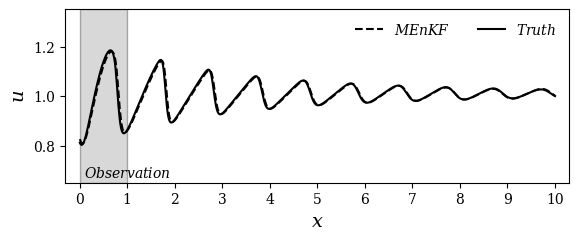}
  \caption{$t=10.60$}
  \end{subfigure}
  \caption{\label{fig:MgENKF_states_rc8} 
Estimations obtained by MEnKF for $r_\text{\tiny C}=8$ at
$t=1$ (a), $t=3.88$ (b)  and $t=10.60$ (c). Times are given in $t_c$ units.
The grey shaded area corresponds to the observation window.  
}
\end{figure}

Results are now investigated for increasing values of $r_\text{\tiny C}$. Results for $r_\text{\tiny C}=8$ are shown in Fig.~\ref{fig:MgENKF_states_rc8}. Minor differences between the state estimation and the true state can be observed in this case. This discrepancy is due to the lack of resolution of the ensemble members. In fact, the resolution in this case is of $10$ mesh elements per characteristic length. This number of points is arguably not enough to provide an accurate representation of the sinusoidal waves which are imposed at the inlet. However, one can see that no spurious numerical effects are observed as the estimator provides a smooth, continuous prediction of the velocity. The discrepancy between the true state and the state estimation is mainly associated with an erroneous calculation of the Kalman gain due to the under-resolution of the ensemble, which also affects the parameter estimation. A combined analysis of Fig.~\ref{fig:amplitude_avg_DENKF} and \ref{fig:MgENKF_states_rc8} shows that, due to the lack of accuracy in the estimation of $\theta_1$, the variable $u$ is over-predicted for $t < 2 t_c$ while it is slightly under-predicted for $t > 4 t_c$.    

\begin{figure}[htbp]
  \begin{subfigure}{\textwidth}
  \includegraphics[width=1\linewidth]{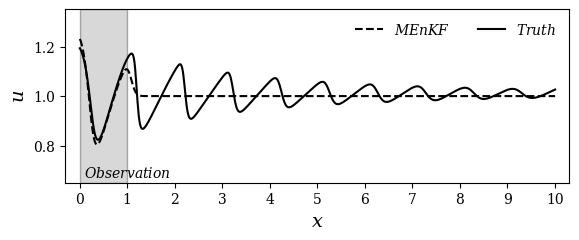}
  \caption{$t=1$}
  \end{subfigure}\par\medskip
  \begin{subfigure}{\textwidth}
  \includegraphics[width=1\linewidth]{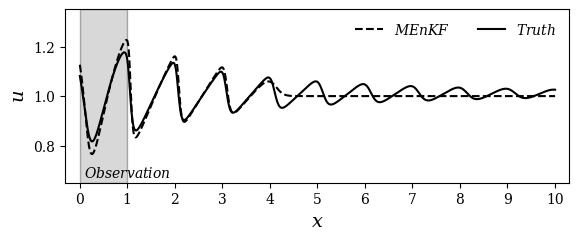}
  \caption{$t=3.88$}
  \end{subfigure}\par\medskip
  \begin{subfigure}{\textwidth}
\includegraphics[width=1\linewidth]{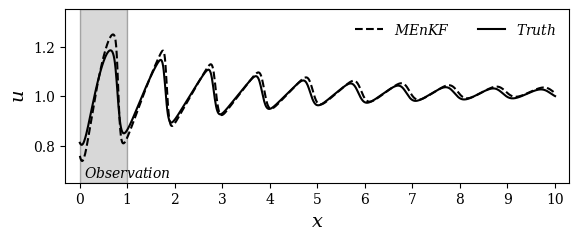}
\caption{$t=10.60$}
  \end{subfigure}
  \caption{\label{fig:MgENKF_states_rc16} 
Estimations obtained by MEnKF for $r_\text{\tiny C}=16$ at
$t=1$ (a), $t=3.88$ (b)  and $t=10.60$ (c). Times are given in $t_c$ units.
The grey shaded area corresponds to the observation window.   
}
\end{figure}

At last, the case for $r_\text{\tiny C}=16$ is shown in Fig.~\ref{fig:MgENKF_states_rc16}. In this case, the mesh elements are only $5$ times smaller than the characteristic length $\lambda$. Despite the important under-resolution of the ensemble members, which severely affects the estimation of $\theta_1$ and $\theta_2$, the state estimation still adequately represents the main features of the dynamical system.

\begin{figure}[htbp]
\centering
\includegraphics[width=1.0\textwidth]{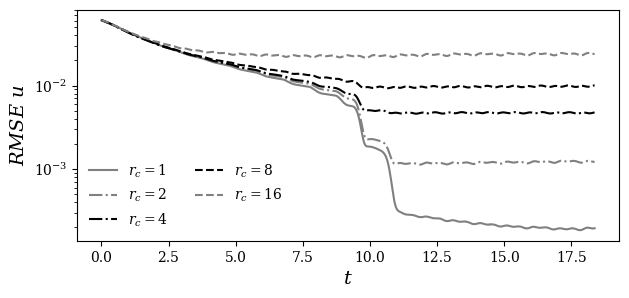}
\caption{\label{fig:stats_DENKF} 
Time evolution of the RMS error of $u$ for $r_\text{\tiny C}=1,2,4,8,16$.
}
\end{figure}

The discrepancy between the truth and the state estimation is measured via the time-dependent relative Root Mean Square Error (RMSE), \textit{i.e.}
\begin{equation}
\text{RMSE}(k)=
\sqrt{\frac{
\displaystyle
\int_{x} \left[\left(u^\text{\tiny F}_k\right)^\text{a}(x) -\left(u^\text{\tiny F}_k\right)^\text{True}(x)\right]^2\diff x
}{
\displaystyle
\int_{x}\left[\left(u^\text{\tiny F}_k\right)^\text{True}(x)\right]^2\diff x}
}
\label{eq:L2norm_u}
\end{equation}
The results are shown in Fig.~\ref{fig:stats_DENKF} for different values of the coarsening ratio $r_\text{\tiny C}$. One can see that the error decreases slowly for $t<10 t_c$. This threshold time corresponds to a complete advective cycle in the physical domain. After this transient, the error may rapidly decrease before reaching a quasi-asymptotic behaviour. 

\begin{figure}[htbp]
\includegraphics[width=1\textwidth]{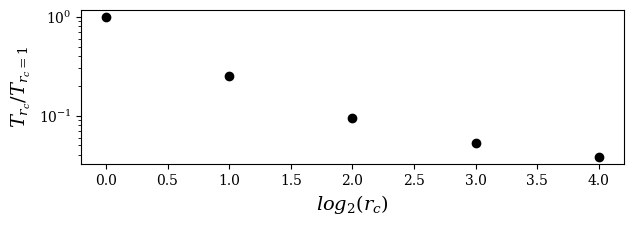}
\caption{\label{fig:calc_time_burgers}
Computational time required to perform a full assimilation cycle for $r_\text{\tiny C}=1,2,4,8,16$. Results are normalized over the computational time required for $r_\text{\tiny C}=1$.
}
\end{figure}

One can also see that, once convergence is reached, the asymptotic RMSE value decreases with lower $r_\text{\tiny C}$ values, as expected. However, lower $r_\text{\tiny C}$ values are as well associated with larger computational costs, so that a trade-off between accuracy and required resources must be found.
This aspect is further investigated considering the computational resources required to perform a full assimilation window for each $r_\text{\tiny C}$ value. In Fig.~\ref{fig:calc_time_burgers}, results are shown and normalized over the case $r_\text{\tiny C}=1$. One can see that the computational resources required rapidly decrease with increasing $r_\text{\tiny C}$, even for this simple one-dimensional test case. For large values of $r_\text{\tiny C}=8,\,16$, one can see that the computational resources reach a plateau. Here the computational time to perform the DA procedures, which is the same for every $r_\text{\tiny C}$, is of similar order of magnitude of the calculation for the time advancement of the ensemble members.

In summary, the present analysis assesses the performance of the MEnKF tool for varying mesh resolution of the ensemble members. As expected, the accuracy of the state and parameter estimations diminishes for increasing $r_\text{\tiny C}$, but so does the computational cost. In addition, it was observed that the accuracy significantly drops when the mesh resolution is not able to provide a suitable description of the main scales characterizing the flow. Such a significant discrepancy for a relatively simple test case stresses how a minimal resolution threshold must be achieved in order to capture the essential physical features and to obtain a successful state estimation. In conclusion, the results of the parametric optimization may become unreliable with extreme under-resolution of the ensemble members. The dispersive and diffusive effects of the coarse grid on the state prediction deserve an in-depth analysis, which is out of the scope of this work. Here, the discretization schemes used are the same for the fine and coarse grids. However, one could argue that choosing adapted schemes for progressively coarser grids could improve the performance of the estimator.

\section{Acoustic propagation of sinusoidal wave}
\label{sec:euler1D}

The MEnKF strategy is now applied to a more complex physical system, namely the inviscid one-dimensional Euler equations:
\begin{eqnarray}
\frac{\partial \rho}{\partial t}+\frac{\partial (\rho u)}{\partial x}= 0\label{eq:e1c} \\
\frac{\partial (\rho u)}{\partial t}+\frac{\partial ((\rho u) u)}{\partial x}+\frac{\partial p}{\partial x}= 0\label{eq:e1m}\\
\frac{\partial \left( \rho E\right)}{\partial t}+\frac{\partial (\left(\rho E\right) u)}{\partial x}+\frac{\partial (pu)}{\partial x}= 0\label{eq:e1e}
\end{eqnarray}
where $\rho$ is the density, $u$ is the velocity, $p$ is the pressure and $E$ is the total energy per unit mass. In this case, viscous effects are absent, but acoustic propagation affects the evolution of the flow. The equations are discretized using the finite difference method. Second-order centered schemes are used for the derivatives in space and a first-order scheme for the time integration to obtain the general form of discretized representation as given by  \eqref{eq:generalDiscretizedFormInt}. A centered sixth-order numerical filter is included to damp numerical spurious oscillations \cite{Bogey_Bailly_2004}. The numerical algorithm is used to analyse the acoustic propagation of a sinusoidal wave with a time-varying amplitude. To do so, a Dirichlet time-varying condition is imposed at the inlet for the velocity:
\begin{equation}
u(x=0,t)=u_0 \left(1+ \theta(t)\sin(\omega t)\right)
\label{eq:inlet-1D-euler}
\end{equation}
$u_0$ is set in order to impose an inlet Mach number $M=\frac{u_0}{a}=0.4$, where $a$ is the speed of the sound. The amplitude of variation in $\theta$ is set at sufficiently low value to allow a flow evolution mainly driven by acoustic phenomena. The inlet velocity perturbation creates an acoustic wave that is transported along the domain with a speed of $u_0+a$. The characteristic velocity and length of the test case are $u_c=u_0+a$ and $\lambda$. Similarly to the analysis in Sec.~\ref{sec:Burgers1D}, $\lambda$ represents the wavelength of the signal imposed at the inlet. The characteristic time of the system is defined as $t_c=\lambda / u_c$. 

The sinusoidal behaviour of the velocity at the inlet is characterized by a constant frequency $f_c=1/t_c$ with $\omega=2\pi f_c$. However, the time-varying amplitude of the sinusoidal wave is driven by the parameter $\theta(t)=\theta_0(1+\sin(\omega_{\theta} t))$, where $\theta_0$, $\omega_{\theta}=2 \pi f_c/b$ and $b$ are constants. The density at the inlet is fixed $\rho(x=0,t)=\rho_0$ as well as the total energy per unit mass $E(x=0,t)=E_0$, which is calculated as $E_0=e+0.5 u_0^2$. The outlet boundary condition is extrapolated from the nearest points to the outlet. The initial condition imposed at  $t=0$ is $u(x,t=0)=u_0$, $\rho(x,t=0)=\rho_0$ and $E(x,t=0)=E_0$ everywhere in the physical domain. The fluid is an ideal gas with $\gamma=1.4$, $\rho_0=1.17$ and $T_0=300$ in S.I. units. Note that from now on, the units
of measure are the characteristic magnitudes.

The computational domain has been set to a size of $L_x=10$. A uniform mesh distribution is used for every calculation. Similarly to the analysis in Sec.~\ref{sec:Burgers1D}, $80$ mesh elements are used to discretize a characteristic length $\lambda$ for a total of $N_x=800$ elements in the domain. Finally, the normalized value of $\Delta t$ is set to $\Delta t=0.0006$.

\begin{figure}[htbp]
\centering
\includegraphics[width=1\textwidth]{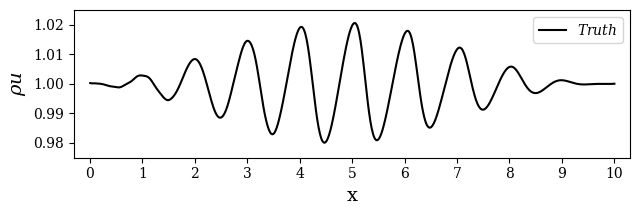}
\caption{\label{fig:1D-Euler-truth}Field $\rho u$ of the one-dimensional inviscid Euler equations at $t=17.3$ for $\theta_0=0.015$ and $\omega_\theta=\omega/10$ (true state).
} 
\end{figure}

A preliminary simulation is performed using an explicit Euler scheme for time derivative. The parameters which describe the amplitude of the velocity oscillation imposed at the inlet are set to $\theta_0=0.015$ (true value) and $b=10$. A flow visualization of the wave patterns is shown in Fig.~\ref{fig:1D-Euler-truth} at $t=17.3$. The fully developed state obtained for $t=10$ is used to initialize a new simulation from $t=0$. This simulation is run for a total time of $T_\text{ref}=110$. As in Sec.~\ref{sec:Burgers1D}, this simulation is used as true state and it is sampled every $30$ time steps to obtain observations in the space region $x \in [0, 1]$. The sampled data, which is used as observation, is artificially perturbed using a Gaussian noise of variance $R=0.09$.    

The DA procedure is identical to the analysis presented in Sec.~\ref{sec:Burgers1D}. The assimilation is composed by the base simulation, which is run on the fine grid, and $N_\text{e}=100$ simulations on a coarse-grid level. In this section, only one coarsening ratio is considered ($r_\text{\tiny C}=4$). The estimator is used to dynamically track the value of the parameter $\theta$, which evolves in time. No a priori knowledge about the behaviour of the parameter is used. This time optimization is much more challenging when compared with the inference of a constant parameter as done in Sec.~\ref{sec:Burgers1D}. A similar analysis using a classical Kalman smoother was recently proposed by Mons et al. \cite{Mons2016_jcp}.

For each coarse grid simulation of the estimator, $\theta$ is initially assumed to be a random Gaussian phenomenon $\theta \sim \mathcal{N}(0, Q_{\theta})$. The initial value of the covariance is $Q_{\theta}(t=0)=6.4\times 10^{-5}$. As shown in the previous case, the value initially imposed at the inlet of the fine-grid simulation is $\theta=0$, while random values are selected for each ensemble member on the coarse grid level following the normal distribution introduced above. The variance of the parameter $\theta$ for the ensemble members is artificially increased, as in the classical Dual EnKF algorithm. As described in \ref{sec:MEnKF}, we add to the estimated parameter of each member of the ensemble a Gaussian noise of diagonal covariance $\mathbf{\Sigma}_k^{\mathbf{\theta}}=10^{-10}$. Extensive numerical tests have been performed and the results show the importance of $\mathbf{\Sigma}_k^{\mathbf{\theta}}$. The value chosen for $\mathbf{\Sigma}_k^{\mathbf{\theta}}$ is arbitrary and works only for this test-case. 
In general, the estimation of $\mathbf{\Sigma}_k^{\mathbf{\theta}}$ is challenging, since a priori information about the dynamic behaviour of $\theta$ is required. 
Nonetheless, $\theta$ can be determined heuristically: the artificial variance added to the parameters, $\mathbf{\Sigma}_k^{\mathbf{\theta}}$, should be of the same order of magnitude as the true rate of change of $\theta$ in a single assimilation cycle. When this parameter is underpredicted, the variance in $\theta$ is too small for the Dual EnKF to perform, while showing a oscillatory behaviour in the parameter estimation if $\mathbf{\Sigma}_k^{\mathbf{\theta}}=10^{-10}$ is overpredicted. 

This study is performed for three different values of $f_a$, the normalized period between successive assimilations defined as $f_a=t_c/t_a$. 
This normalized analysis frequency corresponds to the number of analysis phases per characteristic time of simulation $t_c$. Three different values of $f_a$ are investigated: $f_a=2,\,10,\,55$.  

The estimator is run for a total  simulation time $T_\text{ref}=110$, which encompasses $220$ to $6000$ DA analysis phases, depending on the value of $f_a$. At the end of each analysis, the mean value and the variance of the amplitude $\theta$ are updated following the Dual EnKF technique \cite{DENKF_MORADKHANI2005}, similarly to what was done in Sec.~\ref{sec:Burgers1D}.

\begin{figure}[htbp]
  \begin{subfigure}{\textwidth}
  \includegraphics[width=1\linewidth]{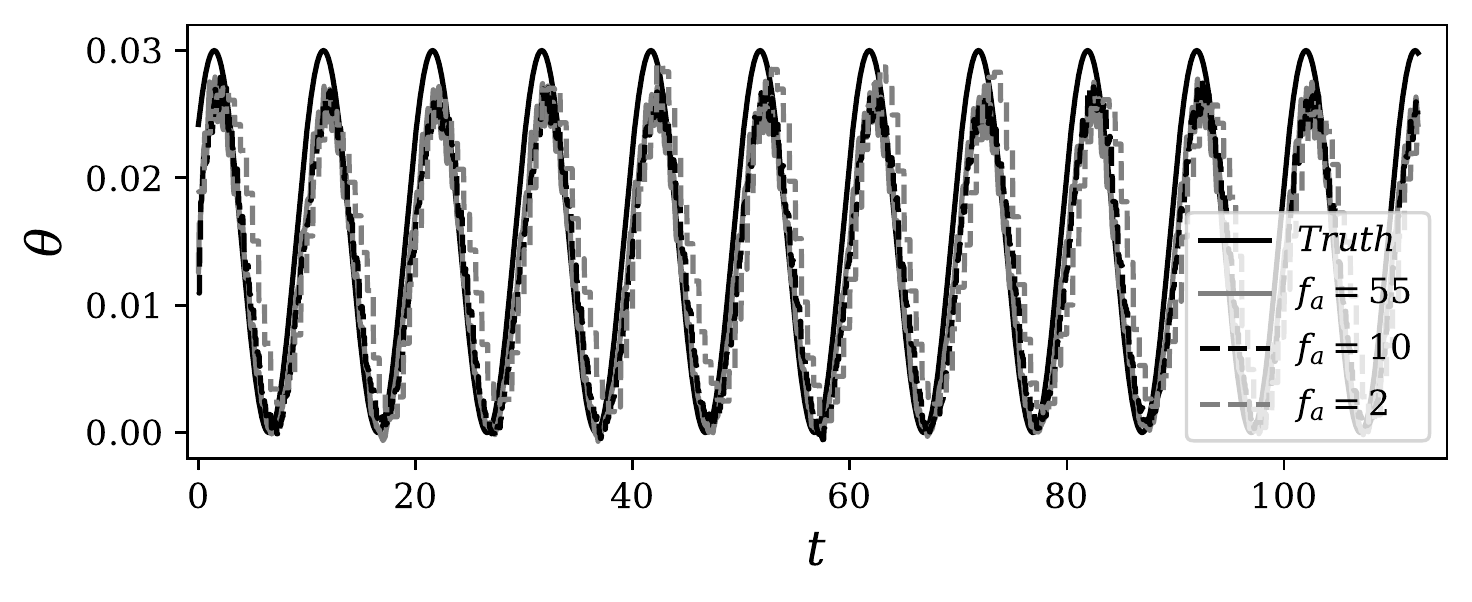}
  \end{subfigure}\par\medskip
  \begin{subfigure}{\textwidth}
  \includegraphics[width=1\linewidth]{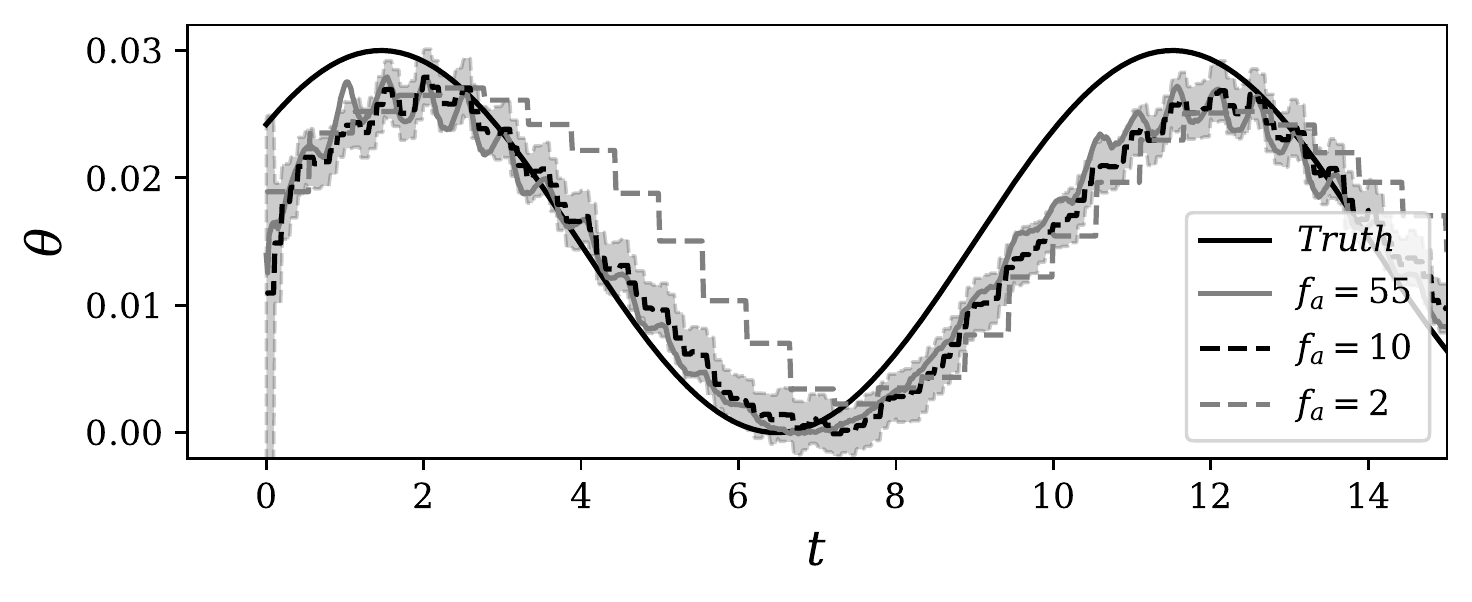}
  \end{subfigure}\par\medskip
  \caption{\label{fig:parameter_euler} Time estimation of the parameter $\theta$ driving the amplitude of the sinusoidal acoustic wave. Results are shown for $f_a=2,\,10,\,55$ and compared to the true value of $\theta$. In the top image, $\theta$ is shown in the whole assimilation window $t \in [0,110]$. In the bottom image, a zoom for $t \in [0,15]$ is shown. The shaded area represents the $95\%$ credible interval for the case $f_a=10$. }
\end{figure}

The results for the estimation of the time-varying parameter $\theta$ are reported in Fig.~\ref{fig:parameter_euler}. 
The time evolution of $\theta$ are correctly estimated for the three values of $f_a$.
This is an important result, considering that no a priori information was provided for the evolution of this parameter. A more detailed analysis reveals a lag in the parameter estimation. The application of a simple Kalman filter seems to be responsible for this result, while a Kalman Smoother (KS) should have been used to obtain a better synchronization. However, considering that the implementation of a KS is straightforward in this case and that observation is always provided close to the inlet, we considered that the increase in computational resources required by the KS were not deserved. We find that the lag increases when a relatively small number of DA analyses is done. One can see that the prediction is significantly degraded for $f_a=2$, while similar results are obtained for $f_a=10,55$. In addition, $\theta$ tends to be underestimated (around $10\%$) when it reaches its maximum value. This result is arguably associated with the under-resolution of the coarse level of the grid, where the gradients of physical variables are calculated with lower accuracy.

\begin{figure}[htbp]
  \begin{subfigure}{\textwidth}
  \includegraphics[width=1\linewidth]{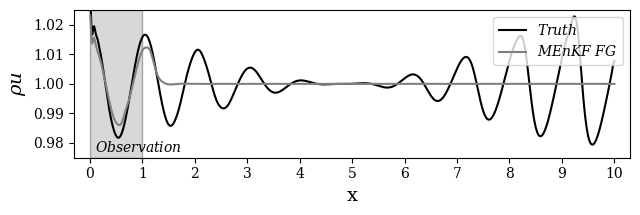}
  \caption{$t=1.23$}
  \end{subfigure}\par\medskip
  \begin{subfigure}{\textwidth}
  \includegraphics[width=1\linewidth]{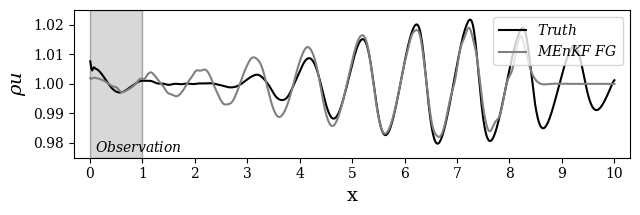}
  \caption{$t=8.32$}
  \end{subfigure}\par\medskip
  \begin{subfigure}{\textwidth}
\includegraphics[width=1\linewidth]{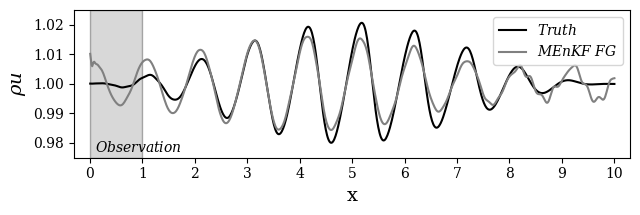}
\caption{$t=16.30$}
  \end{subfigure}
  \caption{\label{fig:MGENKF_states_rc4_euler_f2} 
Estimations by MEnKF of the momentum $\rho u$ normalized by $\rho_0u_0$  for $f_a=2$ at $t=1.23$ (a), $t=8.32$ (b) and $t=16.30$ (c). 
Times are given in $t_c$ units. The grey shaded area corresponds to the observation window.
}
\end{figure}

\begin{figure}[htbp]
  \begin{subfigure}{\textwidth}
  \includegraphics[width=1\linewidth]{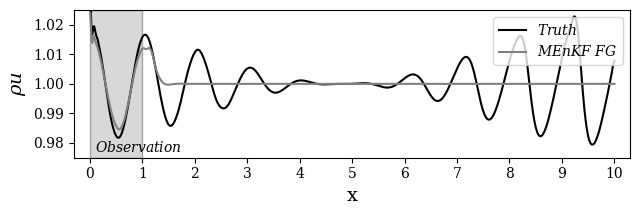}
  \caption{$t=1.23$}
  \end{subfigure}\par\medskip
  \begin{subfigure}{\textwidth}
  \includegraphics[width=1\linewidth]{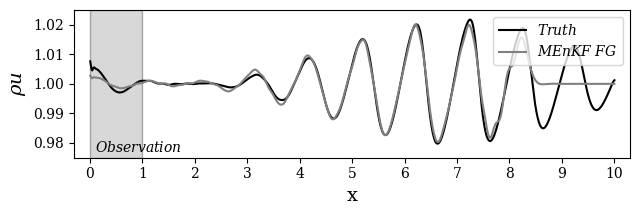}
  \caption{$t=8.32$}
  \end{subfigure}\par\medskip
  \begin{subfigure}{\textwidth}
\includegraphics[width=1\linewidth]{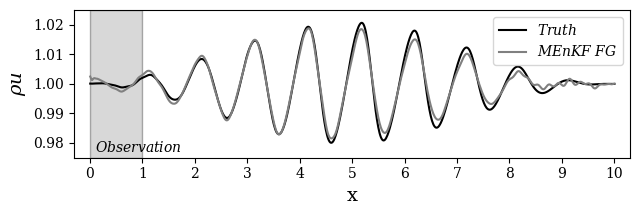}
\caption{$t=16.30$}
  \end{subfigure}
  \caption{\label{fig:MGENKF_states_rc4_euler_f10}
Estimations by MEnKF of the momentum $\rho u$ normalized by $\rho_0u_0$  for $f_a=10$ at $t=1.23$ (a), $t=8.32$ (b) and $t=16.30$ (c). 
Times are given in $t_c$ units. The grey shaded area corresponds to the observation window.  
}
\end{figure}

\begin{figure}[htbp]
  \begin{subfigure}{\textwidth}
  \includegraphics[width=1\linewidth]{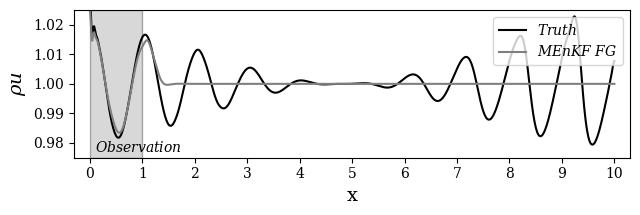}
  \caption{$t=1.23$}
  \end{subfigure}\par\medskip
  \begin{subfigure}{\textwidth}
  \includegraphics[width=1\linewidth]{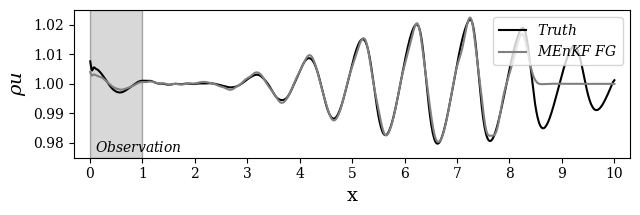}
  \caption{$t=8.32$}
  \end{subfigure}\par\medskip
  \begin{subfigure}{\textwidth}
\includegraphics[width=1\linewidth]{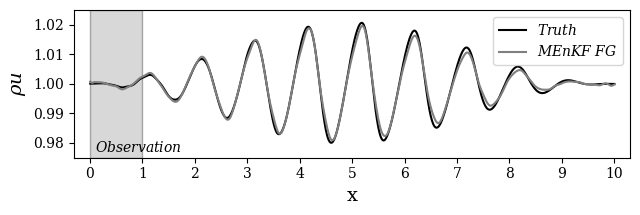}
\caption{$t=16.30$}
  \end{subfigure}
  \caption{\label{fig:MGENKF_states_rc4_euler_f55}
Estimations by MEnKF of the momentum $\rho u$ normalized by $\rho_0u_0$  for $f_a=55$ at $t=1.23$ (a), $t=8.32$ (b) and $t=16.30$ (c). 
Times are given in $t_c$ units. The grey shaded area corresponds to the observation window.  
}
\end{figure}

Now, results dealing with the state estimation are discussed. The predicted physical variable $\rho u$, normalized over the initial value $\rho_0 u_0$, is shown in Fig.~\ref{fig:MGENKF_states_rc4_euler_f2}, \ref{fig:MGENKF_states_rc4_euler_f10} and \ref{fig:MGENKF_states_rc4_euler_f55} for $f_a=2,10$ and $55$, respectively. For $f_a=2$, the state estimation is significantly distant from the truth. It appears that the field correction applied via the Kalman gain is not able to compensate the poor estimation of $\theta$. However, accurate results are observed for $f_a=10$ and $55$. Even though the value of the parameter $\theta$ is not exact, the total state estimation including the correction via Kalman gain is very precise. For the case $f_a=55$, almost no discernible difference is observed between the state estimation and the truth. 

\begin{figure}[htbp]
\centering
\includegraphics[width=1\textwidth]{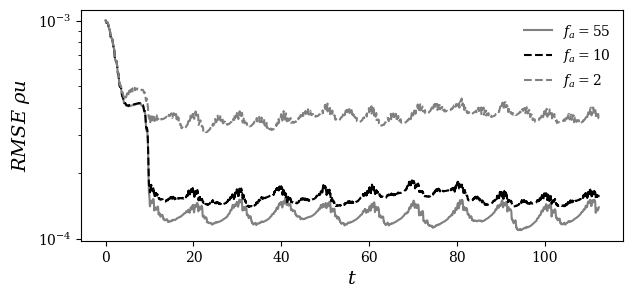}
\caption{\label{fig:RMSE_euler} 
Time evolution of the RMS error of $\rho u$ for $f_a=2,\,10,\,55$.
}
\end{figure}

At last, the relative Root Mean Square Error (RMSE) defined as
\begin{equation}
\text{RMSE}(k)=
\sqrt{\frac{
\displaystyle
\int_{x} \left[\left(\left(\rho u\right)^\text{\tiny F}_k\right)^\text{a}(x)-\left(\left(\rho u\right)^\text{\tiny F}_k\right)^\text{True}(x)\right]^2\diff x
}{
\displaystyle
\int_{x} \left[\left(\left(\rho u\right)^\text{\tiny F}_k\right)^\text{True}(x)\right]^2\diff x}
}
\label{eq:L2norm_u_1D}
\end{equation}
is shown in Fig.~\ref{fig:RMSE_euler}. 
The error achieves a quasi-constant asymptotic behaviour after a complete propagation of the signal in the physical domain ($t \approx 10t_c$). As expected, a low global error is obtained for the cases $f_a=10$ and $f_a=55$. On the other hand, the error for $f_a=2$ case is around $2-3$ times larger. The very small difference in performance between the cases $f_a=10$ and $f_a=55$ indicates that, once a minimal threshold in the assimilation period $t_a$ is reached, the prediction of the estimator exhibits a robust convergence. Similar results were previously observed in three-dimensional simulations \cite{Meldi2018_ftc}. 

\section{Spatially evolving compressible mixing layer}
\label{sec:mixLay2D}

In this section, we consider the compressible Navier-Stokes equations in a two-dimensional physical domain:
\begin{eqnarray}
\label{eq:Mass}
\frac{\partial \rho}{\partial t} &+& div(\rho \mathbf{u}) =0 \\
\label{eq:Momentum}
\dfrac{\partial\left(\rho \mathbf{u}\right)}{\partial t} &+& div(\rho \mathbf{u} \otimes \mathbf{u}) = - \mathbf{grad}\,p + \mathbf{div}\, \overline{\overline{\tau}}  \\
\label{eq:Energy}
\dfrac{\partial\left(\rho E\right)}{\partial t} &+& div (\rho E \mathbf{u}) = - div( p \mathbf{u}) + div (\overline{\overline{\tau}} \mathbf{u}) + div \left(\lambda(T) \mathbf{grad}\,T\right)
\end{eqnarray}
where $\rho$ is the density, $\mathbf{u}$ is the velocity (components $u$ in the streamwise direction and $v$ in the normal direction), $p$ is the pressure, $E$ is the total energy per unit of mass, $\overline{\overline{\tau}}$ is the tensor of the viscous constraints and $T$ is the temperature. To obtain the representation given by \eqref{eq:generalDiscretizedFormInt}, the equations are discretized using the finite difference method. Second-order centered schemes are used for the derivatives in space and a first-order scheme for the time integration. A centered sixth-order numerical filter is included to damp numerical spurious oscillations \cite{Bogey_Bailly_2004}. 

As flow configuration, we consider  the two-dimensional spatially evolving mixing layer at $\text{Re}=100$. For this value of Reynolds number, the flow exhibits unsteady features. It can be shown \cite{Ko2008_pof,Meldi2012_pof} that the characteristics of the mixing layer are strongly affected by the inlet and, in particular, by imposed \textit{ad-hoc} time perturbations. The computational domain has been set to a size of $14\Lambda\times 6\Lambda$ in the streamwise direction $x$ and normal direction $y$, respectively. The characteristic length $\Lambda$, which is taken as reference length  from now on, is given by $\Lambda=A\delta_0$, where $\delta_0$ is the initial vorticity thickness imposed at the inlet. 
The value of the parameter $A$ is determined from the most unstable wavelength determined by Linear Stability Theory (LST). At $\text{Re}=100$, we have 
$A=14.132$. 
The mesh resolution in the horizontal direction is constant for $x \leq 10$.
The size of the elements is $\Delta x = \frac{\delta_0}{8}$. For $x \geq 10$, a sponge zone is established with a coarsening ratio between successive elements which increases from $1.025$ to $1.04$. The resolution in the normal direction is constant and equal to $\Delta y=\frac{\delta_0}{20}$ for $-0.18\leq y \leq 0.18$. Outside this zone, the mesh elements increase in size moving away from the centerline with a constant coarsening ratio of $1.01$.     

The Reynolds number of the flow is calculated as $\text{Re}=(U_1-U_2)\delta_0/\nu=100$ with asymptotic velocities set to $U_1=173.61$ and $U_2=104.17$.
These values correspond to a Mach number $\text{Ma}=0.5$ and $\text{Ma}=0.3$, for each stream, respectively. 
The kinematic viscosity and thermal diffusivity of the flow are considered to be constant and their value is fixed to $\nu=1.568\times 10^{-5}$ and  $\alpha=22.07\times 10^{-7}$, respectively. 
All these quantities are expressed in S.I. units. 
The inlet boundary condition is taken from \cite{Ko2008_pof}. For the velocity field, one has:
\begin{equation}
U_\text{in}=\frac{U_1+U_2}{2}+\frac{U_1-U_2}{2}\tanh\bigg(\frac{2y}{\delta_0}\bigg) + U_\text{pert} \hspace{1cm} -3<y<3\label{eq:NSMXL_UBC}
\end{equation}
\begin{equation}
V_\text{in}=0\label{eq:NSMXL_VBC}
\end{equation}
where $U_\text{in}$ is the streamwise velocity at the inlet and $V_\text{in}$ is the normal velocity. $U_\text{in}$ is estimated as a classical hyperbolic tangent profile plus a time-varying perturbation component:
\begin{equation}
U_\text{pert}= \sum_{i=1}^{N_\text{in}}\epsilon_i\frac{U_1+U_2}{2}[f_i(y)\sin(\omega_it)],\label{eq:NSMXL_UBC_FORCING}
\end{equation}
where $N_\text{in}$ is the total number of perturbation modes and $\epsilon_i$ quantifies the magnitude of each mode. 
The function $f_i(y)=\cos(4n_i\frac{y}{\delta_0})h(y)$ controls the shape of the perturbation of the inlet velocity profile in the normal direction.
The role of $h(y)=1-\tanh^2(\frac{2y}{\delta_0})$ is to damp the perturbation component moving away from the centerline.
The wavelength parameters $n_i$ are tuned according to the LST results. In the following, we consider $N_\text{in}=1$ \textit{i.e.} the inlet perturbation consists of a single mode.
The inlet density is set to be constant so that $\rho_\text{in}=1.177$, as well as the temperature $T=300$ in S.I. units.

\begin{figure}[htbp]
  \begin{subfigure}{\textwidth}
  \includegraphics[width=1\linewidth]{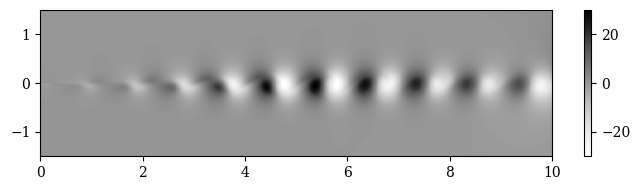}
  \caption{$\epsilon_1$: constant}
  \end{subfigure}\par\medskip
  \begin{subfigure}{\textwidth}
  \includegraphics[width=1\linewidth]{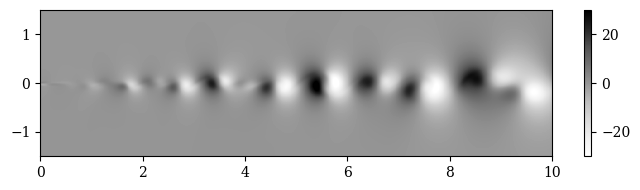}
  \caption{$\epsilon_1$: time-varying}
  \end{subfigure}\par\medskip
  \caption{\label{fig:2D-mxl-trutht10} 
Visualization of the normal momentum $\rho v$ (S.I. units) for the 2D compressible Navier-Stokes equation.
Reference simulation at $t=10$ for a constant value of $\epsilon_1$ $(a)$, and a time-varying value $(b)$.
}
\end{figure}

In this section, the parameters $\mathbf{\theta}$ of the model correspond to $\epsilon_i$, the variable governing the amplitude of the perturbation. Two different scenarios are studied. In the first one, the parameter $\epsilon_i$ is constant.
In the second one, it is a time-dependent coefficient. 
In the first case, the reference simulation is done using a constant single mode for the inlet perturbation $\epsilon_1=\epsilon=\text{cte}$.
In the second case, $\epsilon_1$ varies in time following a sinusoidal form: $\epsilon_1=\epsilon(1+\sin(\omega_{\epsilon}t))$. The values of the numerical parameters characterizing the perturbation are $\epsilon=0.15$, $n_1=0.4\pi$, $\omega_1=1/t_c$ and $\omega_{\epsilon}=0.62\omega_1$, where $t_c=2\Lambda/(U_1+U_2)$ is the average advection time. A flow visualization of $\rho v$ at $t=10$ is shown in Fig.~\ref{fig:2D-mxl-trutht10} for the two cases. 
In both cases, one can clearly observe the emergence of coherent structures.
When a constant value of $\epsilon_1$ is used, the coherent structures are virtually aligned.
When $\epsilon_1$ is time-varying, one can observe the emergence of more complex pairing patterns.

The DA procedure is performed using the following elements:
\begin{itemize}
\item[-] The model is the discretized version of the system given by \eqref{eq:Mass} - \eqref{eq:Energy}. The features of the fine mesh level were previously introduced.
For the coarse grid level, a homogeneous coarsening ratio $r_\text{\tiny C}=4$ is employed. 
The simulation is initialized in the physical domain using a simple hyperbolic tangent profile with no perturbation (see \eqref{eq:NSMXL_UBC}). 
We consider that no prior information is available on the time evolution of $\epsilon_1$.
At $t=0$, this coefficient is fixed to be a random Gaussian phenomenon $\epsilon_1 \sim \mathcal{N}(0, Q_a)$ where the initial value of $Q_a(t=0)=0.0625$. Similarly to the cases analyzed in Sec.~\ref{sec:Burgers1D}, the value imposed on the main fine-grid simulation at $t=0$ is $\epsilon_1=0$, while random values are imposed for each ensemble member on the coarse grid level. The size of the ensemble is $N_\text{e}=100$.
\item[-]  The observation is sampled from the reference simulations shown in Fig.~\ref{fig:2D-mxl-trutht10}, which are run for a total simulation time of $T_\text{ref}=40$ in $t_c$ units. A fully developed state obtained from a prior simulation at $t=10$ is used to initialize the simulations
at $t=0$. 
Data are sampled every $30$ time steps in the region $x \in [0, 0.55]$ and $y \in [-0.16, 0.16]$. 
Considering the results obtained in Sec.~\ref{sec:euler1D}, the update frequency is chosen sufficiently high to assure a good estimation. 
The observations are made from the instantaneous fields $\rho u$ and $\rho v$. 
The data used as observation are artificially perturbed using a Gaussian noise of variance $R=1$.    
\end{itemize}

The estimation algorithm is run over a time window equal to $T_\text{DA}=40$  which encompass roughly $3200$ DA analysis phases. 
This value corresponds to four complete advections in the whole physical domain. At the end of each analysis, the mean value and the variance of the coefficient $\epsilon_1$ are updated following the Dual EnKF technique \cite{DENKF_MORADKHANI2005}, similarly to what was done in Sec.~\ref{sec:Burgers1D}.

\begin{figure}[htbp]
\centering
\includegraphics[width=1\textwidth]{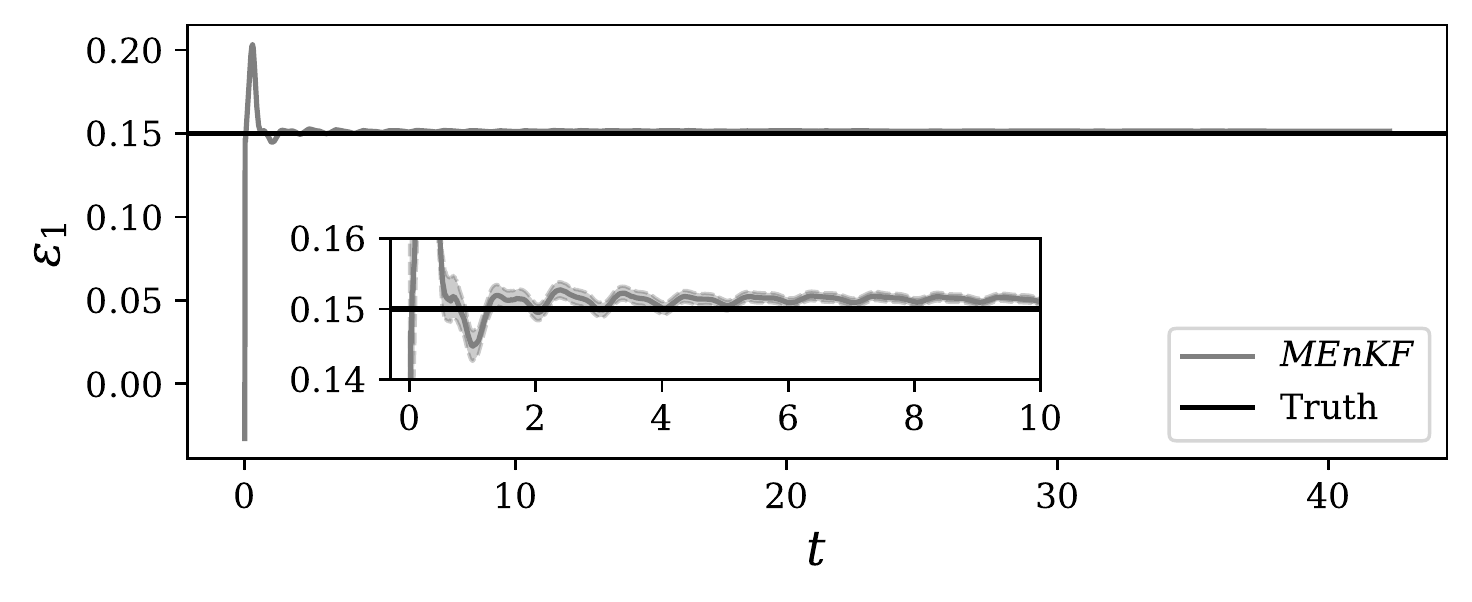}
\caption{\label{fig:param_mxl_c}
Time evolution of the inferred values of $\epsilon_1$. 
The truth corresponds to $\epsilon_1=\text{cte}=0.15$.
In the zoomed region, the shaded area represents the $95\%$ credible interval for the estimated parameter.}
\end{figure}

\begin{figure}[htbp]
  \begin{subfigure}{\textwidth}
  \includegraphics[width=1\linewidth]{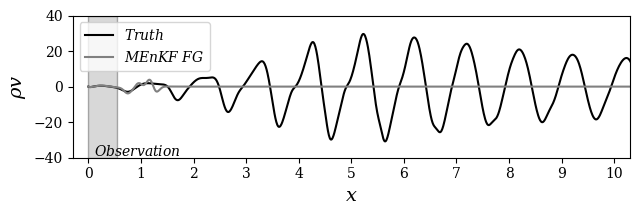}
  \caption{$t=1$}
  \end{subfigure}\par\medskip
  \begin{subfigure}{\textwidth}
  \includegraphics[width=1\linewidth]{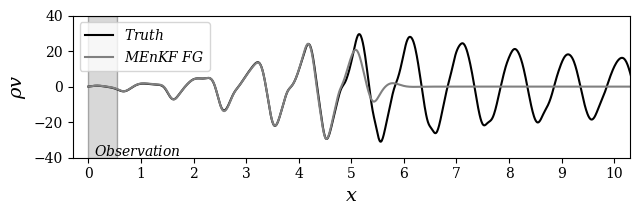}
  \caption{$t=5$}
  \end{subfigure}\par\medskip
  \begin{subfigure}{\textwidth}
\includegraphics[width=1\linewidth]{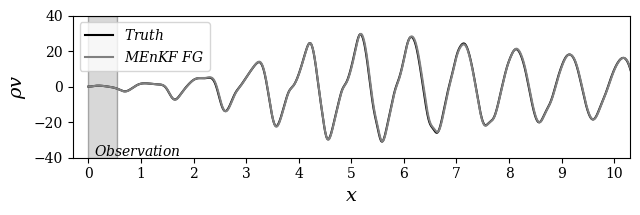}
\caption{$t=30$}
  \end{subfigure}
  \caption{\label{fig:MGENKF_states_rc4_c} 
Estimations obtained by MEnKF of the momentum $\rho v$ (S.I. units) at the centerline $y=0$ of the mixing layer.
Results at $t=1$ (a), $t=5$ (b) and $t=30$ (c) for the case $\epsilon_1=\text{cte}=0.15$.
}
\end{figure}

First, results dealing with the constant value of $\epsilon_1$ are discussed.  
The parameter estimation for this case is almost exact, as observed in Fig.~\ref{fig:param_mxl_c}. The convergence towards the exact value of $\epsilon_1$ is obtained in roughly $0.5$ units of time, after a first transient where a slight over-prediction is observed. When combined with the correction obtained via the Kalman gain, the accuracy of the state estimation is excellent, as shown in Fig.~\ref{fig:MGENKF_states_rc4_c}.

The value of $\rho v$ that is predicted at the centerline of the mixing layer is shown in Fig.~\ref{fig:MGENKF_states_rc4_c}. At $t=1$, the flow is strongly affected by the KF corrections. After five characteristic times, the flow is perfectly matching the true state upstream. Finally, after the initial transient is dissipated, we observe that the state estimation almost perfectly matches the true state in the whole physical domain.

\begin{figure}[htbp]
  \begin{subfigure}{\textwidth}
  \includegraphics[width=1\linewidth]{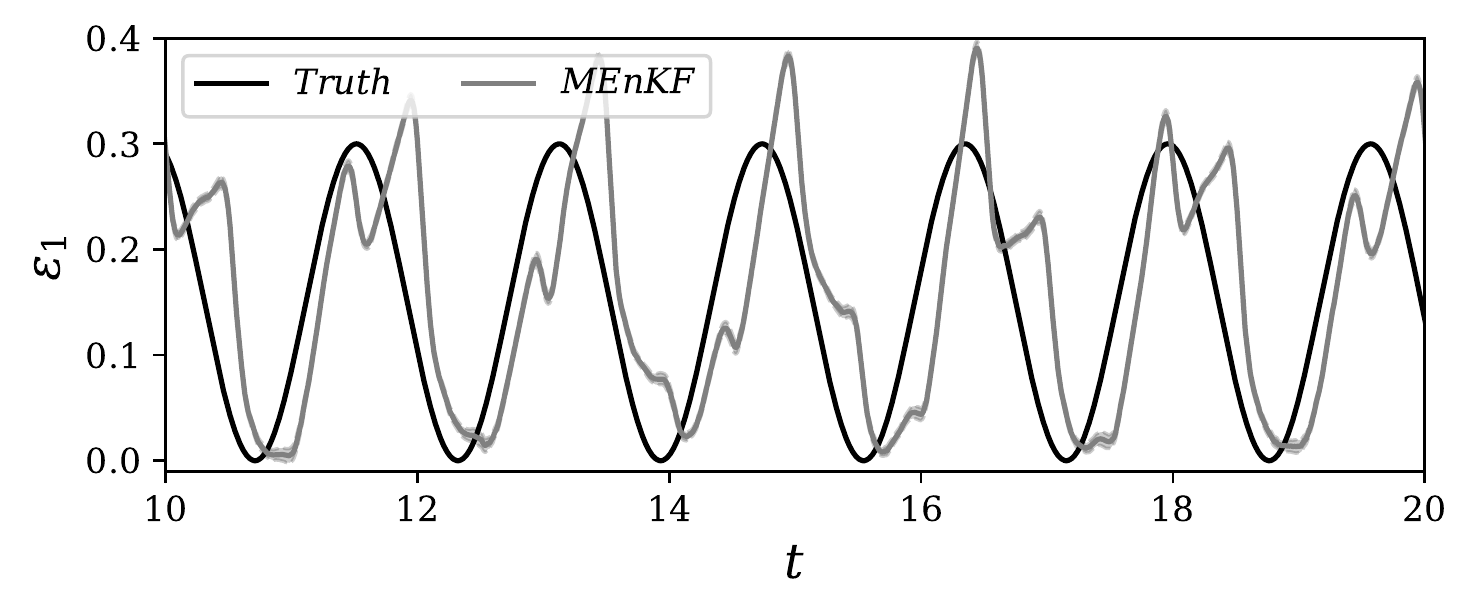}
  \caption{Estimation history}
  \end{subfigure}\par\medskip
  \begin{subfigure}{\textwidth}
  \includegraphics[width=1\linewidth]{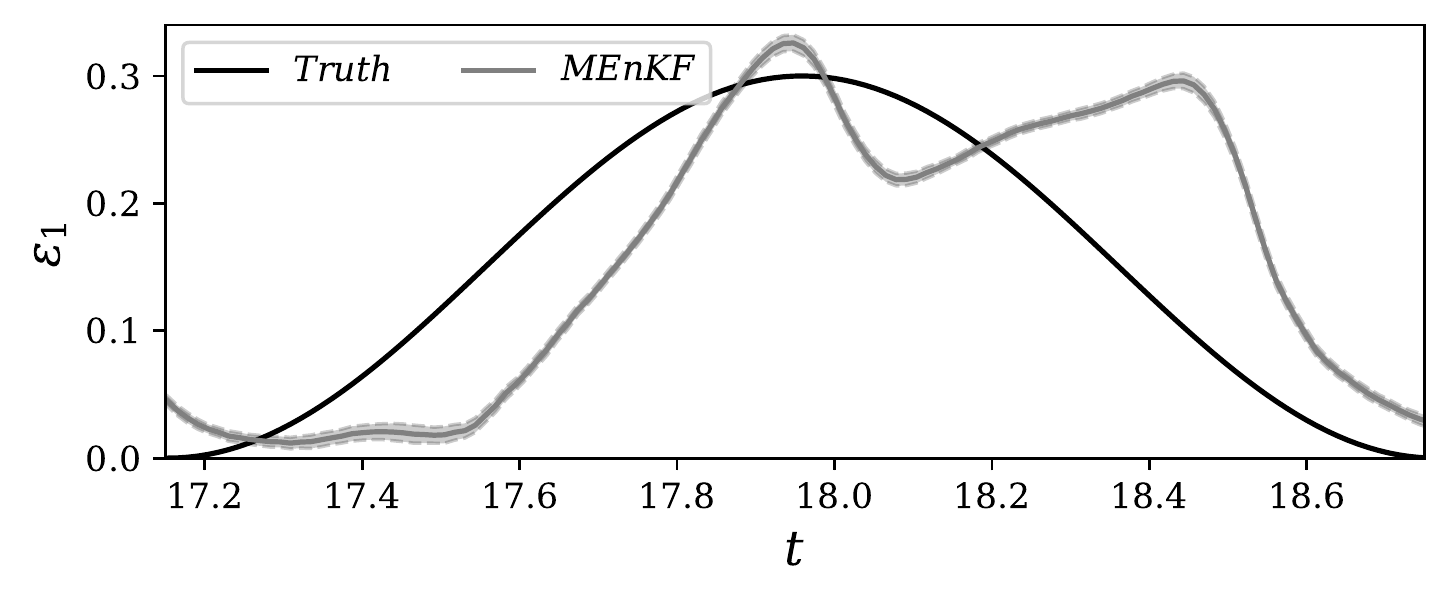}
  \caption{Zoom}
  \end{subfigure}\par\medskip
  \caption{\label{fig:param_mxl_v} 
Time evolution of the inferred values of $\epsilon_1$ for the time-varying reference case. 
(a) Large time window.
(b) Zoomed region. The shaded area represents the $95\%$ credible interval for the estimated parameter.
}
\end{figure}

The results dealing with the time-varying parameter $\epsilon_1$ are now discussed. 
The time evolution of the estimated value of $\epsilon_1$ is reported in Fig.~\ref{fig:param_mxl_v}. 
The overall sinusoidal trend is generally respected, although a relatively small phase lag is visible. This lag does not appear to be larger than the one previously observed for the one-dimensional case based on the Euler equation.
The presence of this delay has probably the same reasons as in Sec.~\ref{sec:euler1D}.
However, in this case, some over prediction of the parameter is locally observed in time, which was not obtained for the wave propagation test case.  

\begin{figure}[htbp]
  \begin{subfigure}{\textwidth}
  \includegraphics[width=1\linewidth]{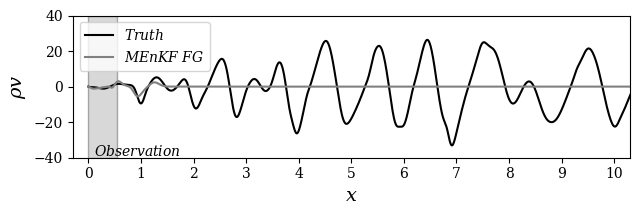}
  \caption{$t=1$}
  \end{subfigure}\par\medskip
  \begin{subfigure}{\textwidth}
  \includegraphics[width=1\linewidth]{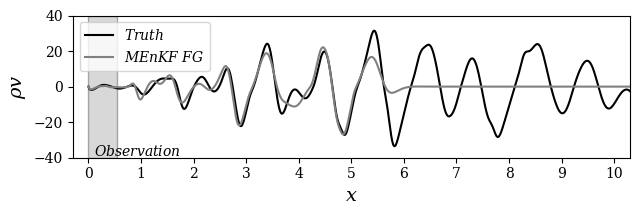}
  \caption{$t=5$}
  \end{subfigure}\par\medskip
  \begin{subfigure}{\textwidth}
\includegraphics[width=1\linewidth]{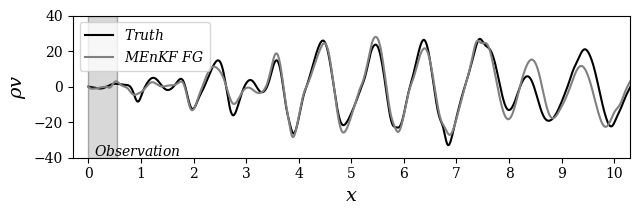}
\caption{$t=30$}
  \end{subfigure}
  \caption{\label{fig:MGENKF_states_rc4_1}
Estimations obtained by MEnKF of the momentum $\rho v$ (S.I. units) at the centerline $y=0$ of the mixing layer.
Results at $t=1$ (a), $t=5$ (b) and $t=30$ (c) for the time-varying $\epsilon_1$.  
}
\end{figure}

\begin{figure}[htbp]
  \begin{subfigure}{\textwidth}
  \includegraphics[width=1\linewidth]{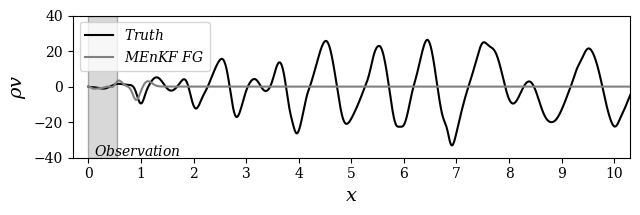}
  \caption{$t=1$}
  \end{subfigure}\par\medskip
  \begin{subfigure}{\textwidth}
  \includegraphics[width=1\linewidth]{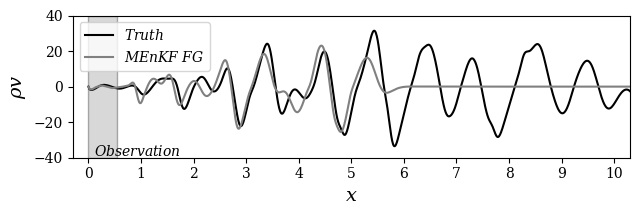}
  \caption{$t=5$}
  \end{subfigure}\par\medskip
  \begin{subfigure}{\textwidth}
\includegraphics[width=1\linewidth]{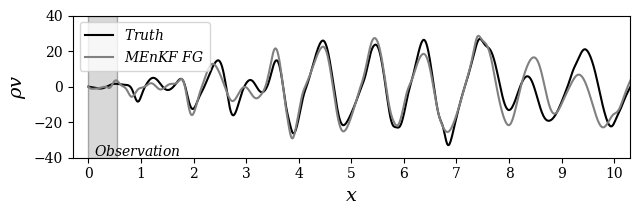}
\caption{$t=30$}
  \end{subfigure}
  \caption{\label{fig:MGENKF_states_rc4_w}
Estimations obtained by MEnKF of the momentum $\rho v$ (S.I. units) at the centerline $y=0$ of the mixing layer.
Here, MEnKF is only used to provide the estimation of $\epsilon_1$.
Results at $t=1$ (a), $t=5$ (b) and $t=30$ (c) for the time-varying $\epsilon_1$.  
}
\end{figure}

The results obtained for the prediction of the normal momentum $\rho v$ are shown in Fig.~\ref{fig:MGENKF_states_rc4_1}. 
One can see that the combination of parameter and state estimations produces an accurate prediction of the flow. Minor differences are observed with the true state.
In particular, the momentum $\rho v$ does not exhibit spurious oscillations which could stem from the field correction determined via the Kalman gain. 
In order to evaluate the respective influence of the parameter estimation step and state estimation phase, a test case is run in which only the parameter estimation is performed. 
That is, the state estimation obtained on the coarse-grid level is not included in the steps $4$ and $5$ of the algorithm presented in Sec.~\ref{sec:multigrid-EnKF}. 
While the results of the parameter estimation are the same for the two cases, one can see in Fig.~\ref{fig:MGENKF_states_rc4_w} that the prediction is sensibly deteriorated.
 

This observation is quantified by the evaluation of the relative Root Mean Square Error (RMSE), defined as:
\begin{equation}
\text{RMSE}(k)=
\sqrt{\frac{
\displaystyle
\int_{x} \left[\left(\left(\rho v\right)^\text{\tiny F}_k\right)^\text{a}(x)-\left(\left(\rho v\right)^\text{\tiny F}_k\right)^\text{True}(x)\right]^2\diff x
}{
\displaystyle
\int_{x} \left[\left(\left(\rho v\right)^\text{\tiny F}_k\right)^\text{True}(x)\right]^2\diff x}
}
\label{eq:L2norm_u_2D}
\end{equation}

\begin{figure}[htbp]
\centering
\includegraphics[width=1\textwidth]{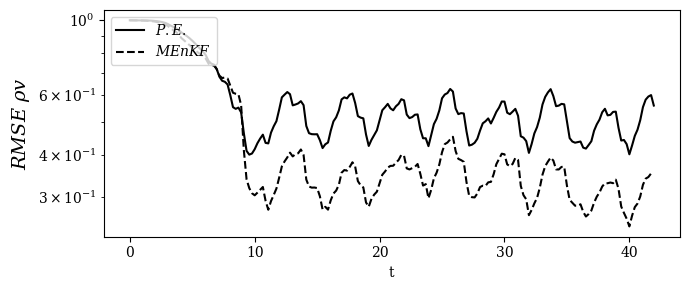}
\caption{\label{fig:stats_mxl_v}
Time evolution of the RMS error of $\rho v$ for the case of a time-varying inlet parameter $\epsilon_1$.
The symbol P.E. corresponds to the case where MEnKF is only used for the estimation of $\epsilon_1$.
The notation MEnKF corresponds to the standard version of the algorithm, including parameter estimation and physical state correction via Kalman gain.
}
\end{figure}

The results, which are shown in Fig.~\ref{fig:stats_mxl_v}, indicate that the accuracy of the complete algorithm is higher when compared to the case in which only the parameter estimation is performed. Therefore, the two operations concurrently provide an improvement in the prediction of the flow.  

At last, an analysis of the conservativity of the algorithm is performed. As previously discussed, the state estimation obtained via EnKF does not necessarily comply with the dynamical equations of the model. This drawback can be responsible for discontinuities in the physical field, which can significantly affect the accuracy and stability of the global algorithm. The analysis is performed considering an indicator $\Gamma_k^\text{\tiny F}$ which measures the conservation of the transversal momentum equation \eqref{eq:Momentum} in discretized form:

\begin{equation}
\frac{\left(\left(\rho v\right)_k^\text{\tiny F}\right)^\text{a}-\left(\left(\rho v\right)_{k-1}^\text{\tiny F}\right)^\text{a}}{\Delta t}-\mathcal{F}_{\rho v}\left(\rho_{k}, \mathbf{u}_{k}, p_{k}, \overline{\overline{\tau}}_{k}\right)=\Gamma_k^\text{\tiny F},
\label{eq:Conservativity_rhov}
\end{equation}
where $\mathcal{F}_{\rho v}$ represents the spatial discretization terms in the transversal momentum equation. In the forecast step performed via the model, $\Gamma_k^\text{\tiny F} = 0$ down to a convergence rate $\delta$ which is prescribed. However, the value of $\Gamma_k^\text{\tiny F}$, at the end of a time step where a forecast-analysis is performed, is strictly connected with the computational strategy employed. Here, three cases are considered for $\epsilon_1=\text{cte}=0.15$:
\begin{itemize}
\item[-] A classical Dual EnKF is performed on the coarse grid and a fine-grid correction is obtained through the ensemble statistics. In this scenario, the state estimation obtained in the step $4$ of the MEnKF algorithm presented in Sec.~ \ref{sec:multigrid-EnKF} is directly projected in the fine mesh space and used as final solution. The step $5$ of the algorithm is not performed.
\item[-] A standard MEnKF algorithm, as described in Sec.~\ref{sec:multigrid-EnKF}.
\item[-] A MEnKF algorithm where the ensemble prediction is  just used to estimate the unknown parameter of the system. No update of the physical solution is performed using the correction via Kalman gain.
\end{itemize}

\begin{figure}[htbp]
  \begin{subfigure}{\textwidth}
  \includegraphics[width=1\linewidth]{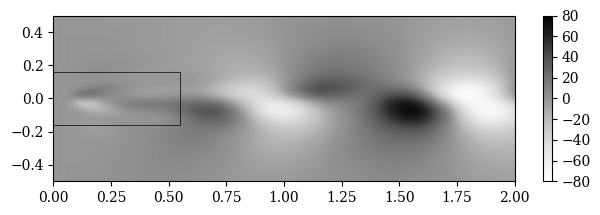}
  \caption{Standard Dual EnKF on the coarse mesh.}
  \end{subfigure}\par\medskip
  \begin{subfigure}{\textwidth}
  \includegraphics[width=1\linewidth]{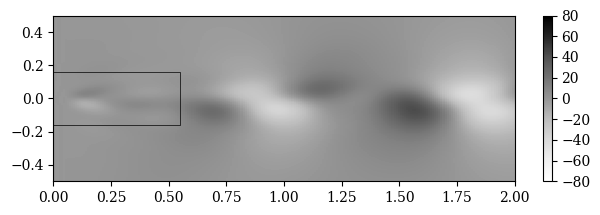}
  \caption{Standard MEnKF.}
  \end{subfigure}\par\medskip
  \begin{subfigure}{\textwidth}
\includegraphics[width=1\linewidth]{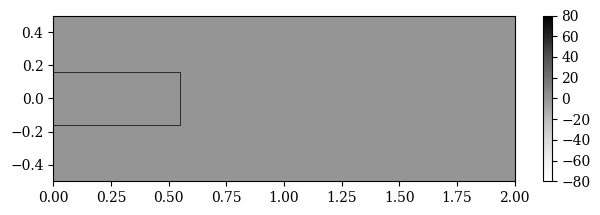}
\caption{MEnKF used only for parameter estimation.}
  \end{subfigure}
  \caption{\label{fig:GAMMA} 
  Analysis of the conservativity of the dynamical model via the normalized quantity $\left(\Gamma_k^\text{\tiny F}\right)^*$. 
  Results are shown for three scenarios: (a) the classical Dual EnKF, (b) the classical MEnKF algorithm and (c) the MEnKF only used for parameter estimation.
  }
\end{figure}

The results are shown in Fig.~\ref{fig:GAMMA} after the first forecast / analysis step. For clarity, we introduce a normalized criterion $\left(\Gamma_k^\text{\tiny F}\right)^*=\frac{\Gamma_k^\text{\tiny F}}{C}$ where $C$ is defined as $\max_k \left\lvert \frac{\left(\left(\rho v\right)_k^\text{\tiny F}\right)^\text{a}-\left(\left(\rho v\right)_{k-1}^\text{\tiny F}\right)^\text{a}}{\Delta t}\right\lvert$.
As expected, $\Gamma_k^\text{\tiny F}=0$ everywhere when MEnKF is only used for the parameter estimation. 
Here, the time advancement of the solution is performed using the model only, which exactly complies with the discretized equation and respects conservativity (up to a convergence error which is negligible). On the other hand, results in Fig.~\ref{fig:GAMMA} (a) show some lack of conservativity in the physical domain for the first scenario. This is also expected, since no constraint is imposed to force the Kalman gain correction to comply with the dynamical  equations. 
Finally, results for the MEnKF are shown in Fig.~\ref{fig:GAMMA} (b). The evolution of $\left(\Gamma_k^\text{\tiny F}\right)^*$ is very similar to the results observed for the first scenario. 
However, one can clearly see that this field appears to be sensibly smoothed out by the multigrid iterative procedures in step 4 and 5 of the MEnKF algorithm. As previously discussed, complete conservativity starting from an erroneous state at $k-1$ is possibly not an optimal objective, while one wants a regularized solution to avoid affecting the precision of the global calculation. On this last objective, the MEnKF appears to provide a better result when compared with the classical Dual EnKF, described in the first scenario. 
Considering also that the MEnKF showed better accuracy than the algorithm relying on parameter estimation only, one can conclude that the MEnKF provides an efficient compromise between global accuracy and regularization of the solution. 
In order to draw more information about this important aspect, the MEnKF algorithm needs to be tested for the simulation of three-dimensional compressible flows, where the Kalman gain correction may be responsible for important acoustic phenomena which are not observable in 2D and 1D dynamical systems.

\section{Conclusions}
\label{sec:conclusions}
A sequential estimator based on a Kalman filter approach for Data Assimilation of fluid flows is presented in this research work. This estimator exploits iterative features which are employed in several CFD codes for the resolution of complex applications in fluid mechanics. More precisely, the multilevel resolution associated with the multigrid iterative approach for time advancement is used to generate several low-resolution numerical simulations. These results are then employed as ensemble members to determine i) the correction via Kalman filter, which is then projected on the high-resolution grid to correct a single simulation which corresponds to the numerical model and ii) an optimization of the free parameters driving the simulation. One of the main advantages of the model is that, owing to the iterative procedure for the calculation of the flow variables, the final solution is regularized.

The method, which is referred to as Multigrid Ensemble Kalman Filter (MEnKF), is assessed via the analysis of one-dimensional and two-dimensional test cases and using different dynamic equations. First, the one-dimensional Burgers equation for $Re=200$ is analyzed. Here, the performance of MEnKF is assessed considering several coarsening ratios $r_\text{\tiny C}$, which determines the difference in resolution between the main simulation and the ensemble members. The results for $r_\text{\tiny C}=1$ (i.e. method equivalent to a EnKF) indicate that the State Estimation and the parametric optimization of the inlet provide very high accuracy in the results. With increasing coarsening ratios the quality of the results is progressively degraded, but the main features of the flow are obtained even for very under-resolved ensemble members. In addition, higher $r_\text{\tiny C}$ values are associated with significantly decreased computational costs, so that this method exhibit a potential to be explored for efficient trade off between accuracy and resources required. 

Then, MEnKF is used to track the time evolution of a free parameter for the case of a wave propagation, using a one-dimensional Euler model. Three cases are here investigated, varying the time window between successive assimilations. The estimator can efficiently represent the evolution in time of the parameter, as well as to provide an accurate state estimation. However, the global prediction is significantly degraded if the assimilation window is larger than a threshold value, which is arguably connected to the physical features of the flow. 

At last, the analysis of the two-dimensional spatially evolving mixing layer for $\text{Re}=100$ is performed. The algorithm appears to be well suited for the analysis of unsteady phenomena, in particular for the analysis of time varying free parameters of the simulation. These features are promising for potential application to in-streaming Data Assimilation techniques.  

Future research over this method will target improvement of the projector operators between coarse and fine grid level, in particular for the state matrix $\Psi$. Preliminary tests have shown that, in the case of non-linear models, information from the main, refined simulation can be projected on the coarse mesh level to improve the predictive capabilities of the ensemble members. This implies a more accurate prediction of the state estimation and of the parameter optimization on the coarse level, from which the main refined simulation will benefit. This loop has the potential to improve even more the performance of the MEnKF model, and strategies for efficient application are currently under investigation. Moreover, we plan to test the MEnKF algorithm to more complex configurations involving complex geometries and more challenging parametric optimization problems. The compressible effects are relatively low for the tests performed so far. Further research is planned on test-cases where the compressible effects are more accentuated.

Acknowledgements: Our research activities are supported by the Direction Générale de l'Armement (DGA) and the Région Nouvelle Aquitaine. Prof. Heng Xiao is warmly acknowledged for valuable discussion on the subject.\\

Conflict of interest: The Authors have no conflict of interests.

\appendix

\section{Data Assimilation algorithms}
\label{sec:DA_algorithms}
\graphicspath{{Cordier_Figs/}}
\subsection{Kalman filter algorithm}
\label{sec:KFalgo}
\begin{figure}[htbp]
\centering
\includegraphics[width=1\textwidth]{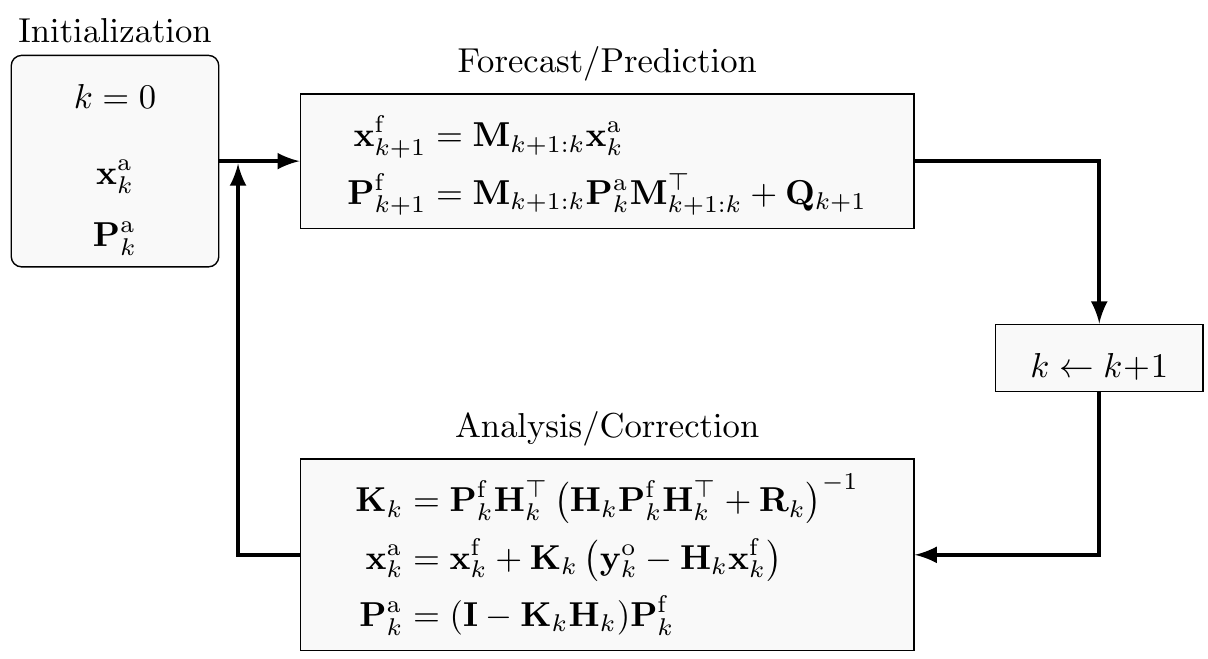}
\caption{\label{fig:KF_Init_analyse}Kalman Filter algorithm. The initialization is made with the analysed state.}
\end{figure}

The Kalman filter algorithm given in Sec.~\ref{sec:KF} corresponds to Fig.~\ref{fig:KF_Init_analyse}.

\subsection{Ensemble Kalman filter algorithm}
\label{sec:EnKFalgo}
An efficient implementation of the EnKF relying on anomaly matrices is given in Algo.~\ref{alg:Stoch_EnKF_Anomaly_Carrassi}. We have used the secant method described in \cite{Asch2016_SIAM} to change the definition of the variable $\mathbf{Y}_k^\text{f}$.

\newpage
\begin{algorithm}
\DontPrintSemicolon
\KwIn{For $k=0,\dots,K$: 
the forward models $\mathbf{\mathcal{M}}_{k:k-1}$, 
the observation models $\mathbf{\mathcal{H}}_k$, 
the observation error covariance matrices $\mathbf{R}_k$}
\KwOut{$\{\mathbf{x}_{k}^{\text{a},(i)}\}$\,;\, $k=0,\cdots,K$\,;\, $i=1,\cdots,N_\text{e}$}
\Begin{
\nlset{1:}Initialize the ensemble of forecasts $\{\mathbf{x}_{0}^{\text{f},(i)}\}$\,;\, $i=1,\cdots,N_\text{e}$\;
\For{$k=0,\dots,K$}{
\nlset{2:}Draw a statistically consistent observation set\,;\, $i=1,\cdots,N_\text{e}$\\
$\mathbf{y}_k^{\text{o},(i)}=\mathbf{y}_k^\text{o} + \mathbf{\epsilon}_k^{\text{o},(i)}\quad\text{with}\quad\mathbf{\epsilon}_k^{\text{o},(i)}\sim \mathcal{N}(0,\mathbf{R}_k)$\;
\nlset{3:}Compute the model counterparts of the observation set\,;\, $i=1,\cdots,N_\text{e}$\\
$\mathbf{y}_{k}^{\text{f},(i)}=\mathbf{\mathcal{H}}_k\left(\mathbf{x}_{k}^{\text{f},(i)}\right)$
\;
\nlset{4:}Compute the ensemble means\\
$\displaystyle\overline{\mathbf{x}_{k}^{\text{f}}}=\frac{1}{N_\text{e}}\sum_{i=1}^{N_\text{e}}\mathbf{x}_{k}^{\text{f},(i)}$\,;\,
$\displaystyle\overline{\mathbf{y}_{k}^{\text{f}}}=\frac{1}{N_\text{e}}\sum_{i=1}^{N_\text{e}}\mathbf{y}_{k}^{\text{f},(i)}$\,;\,
$\displaystyle\overline{\mathbf{\epsilon}_{k}^{\text{o}}}=\frac{1}{N_\text{e}}\sum_{i=1}^{N_\text{e}}\mathbf{\epsilon}_k^{\text{o},(i)}$
\;
\nlset{5:}Compute the normalized anomalies\,;\, $i=1,\cdots,N_\text{e}$\\
$$
\left[\mathbf{X}_k^\text{f}\right]_{:,i}=
\frac{\mathbf{x}_{k}^{\text{f},(i)}-\overline{\mathbf{x}_{k}^{\text{f}}}}{\sqrt{N_\text{e}-1}}
\,;\,
\left[\mathbf{Y}_k^\text{f}\right]_{:,i}=
\frac{%
\mathbf{y}_{k}^{\text{f},(i)}-
\overline{\mathbf{y}_{k}^{\text{f}}}
}{\sqrt{N_\text{e}-1}}
\,;\,
\left[\mathbf{E}_k^\text{o}\right]_{:,i}=
\frac{%
\mathbf{\epsilon}_k^{\text{o},(i)}-\overline{\mathbf{\epsilon}_{k}^{\text{o}}}
}{\sqrt{N_\text{e}-1}}
$$
\;
\nlset{6:}Compute the Kalman gain\\
$\mathbf{K}_k^\text{e}=
\mathbf{X}_k^\text{f}
\left(\mathbf{Y}_k^\text{f}\right)^\top
\left(
\mathbf{Y}_k^\text{f}
\left(\mathbf{Y}_k^\text{f}\right)^\top
+
\mathbf{E}_k^\text{o} \left(\mathbf{E}_k^\text{o}\right)^\top
\right)^{-1}$
\;
\nlset{7:}Update the ensemble\,;\, $i=1,\cdots,N_\text{e}$\\
$
\mathbf{x}_{k}^{\text{a},(i)}=
\mathbf{x}_{k}^{\text{f},(i)}+
\mathbf{K}_k^\text{e}
\left(
\mathbf{y}_k^{\text{o},(i)}-\mathbf{y}_{k}^{\text{f},(i)}
\right)
$
\;
\nlset{8:}Compute the ensemble forecast\,;\, $i=1,\cdots,N_\text{e}$\\
$\mathbf{x}_{k+1}^{\text{f},(i)}=\mathbf{\mathcal{M}}_{k+1:k}(\mathbf{x}_{k}^{\text{a},(i)})$
\;

}
}
\caption{\label{alg:Stoch_EnKF_Anomaly_Carrassi}Stochastic Ensemble Kalman Filter (slightly adapted from \cite{Asch2016_SIAM}). Use of anomaly matrices with $\mathbf{Y}_k^\text{f}=\mathbf{H}_k\mathbf{X}_k^\text{f}$.}
\end{algorithm}

\subsection{Dual Ensemble Kalman filter algorithm}
\label{sec:DualEnKFalgo}
An efficient implementation of the Dual EnKF relying on anomaly matrices is given in Algo.~\ref{alg:Stoch_Dual_EnKF_Moradkhani_Carrassi}. We have slightly adapted this algorithm from \cite{DENKF_MORADKHANI2005}. 

\begin{algorithm}
\DontPrintSemicolon
\KwIn{For $k=1,\dots,K$: 
the forward models $\mathbf{\mathcal{M}}_{k:k-1}$, 
the observation models $\mathbf{\mathcal{H}}_k$, 
the observation error covariance matrices $\mathbf{R}_k$}
\KwOut{$\{\mathbf{\theta}_{k}^{\text{a},(i)}\}$ and $\{\mathbf{x}_{k}^{\text{a},(i)}\}$\,;\, $k=0,\cdots,K$}
\Begin{
\nlset{1:}Initialize $\{\mathbf{\theta}_{0}^{\text{a},(i)}\}$ and $\{\mathbf{x}_{0}^{\text{a},(i)}\}$\;
\For{$k=1,\dots,K$}{
\nlset{2:}Observation ensemble:
\vspace*{-4mm}
\begin{align*}
\mathbf{y}_k^{\text{o},(i)} & =\mathbf{y}_k^\text{o} + \mathbf{\epsilon}_k^{\text{o},(i)}\quad\text{with}\quad\mathbf{\epsilon}_k^{\text{o},(i)}\sim \mathcal{N}(0,\mathbf{R}_k)\\
\displaystyle
\mathbf{R}_k^\text{e}
& =
\frac{1}{N_\text{e}-1}
\sum_{i=1}^{N_\text{e}}
\mathbf{\epsilon}_k^{\text{o},(i)}
\left(
\mathbf{\epsilon}_k^{\text{o},(i)}
\right)^\top
\end{align*}
\;
\vspace*{-9mm}
\nlset{3:}Parameter forecast:
\vspace*{-4mm}
\begin{align*}
\mathbf{\theta}_{k}^{\text{f},(i)} & = \mathbf{\theta}_{k-1}^{\text{a},(i)}+\mathbf{\tau}_{k}^{(i)}\quad\text{with}\quad\mathbf{\tau}_{k}^{(i)}\sim \mathcal{N}(0,\mathbf{\Sigma}_k^\mathbf{\theta})\\
\mathbf{x}_{k}^{\text{f},(i)} & = \mathbf{\mathcal{M}}_{k:k-1}(\mathbf{x}_{k-1}^{\text{a},(i)},\mathbf{\theta}_{k}^{\text{f},(i)})\\
\mathbf{y}_{k}^{\text{f},(i)} & = \mathbf{\mathcal{H}}_k\left(\mathbf{x}_{k}^{\text{f},(i)}\right)
\end{align*}
\;
\vspace*{-9mm}
\nlset{4:}Compute the normalized anomalies
\vspace*{-4mm}
$$
\left[\mathbf{\Theta}_k^\text{f}\right]_{:,i}=
\frac{\mathbf{\theta}_{k}^{\text{f},(i)}-\overline{\mathbf{\theta}_{k}^{\text{f}}}}{\sqrt{N_\text{e}-1}}
\: 
;
\:
\left[\mathbf{Y}_k^\text{f}\right]_{:,i}=
\frac{%
\mathbf{y}_{k}^{\text{f},(i)}-
\overline{\mathbf{y}_{k}^{\text{f}}}
}{\sqrt{N_\text{e}-1}}
\: 
;
\:
\left[\mathbf{E}_k^\text{o}\right]_{:,i}=
\frac{%
\mathbf{\epsilon}_k^{\text{o},(i)}-\overline{\mathbf{\epsilon}_{k}^{\text{o}}}
}{\sqrt{N_\text{e}-1}}
$$
\;
\vspace*{-9mm}
\nlset{5:}Parameter update:
\vspace*{-4mm}
\begin{align*}
\mathbf{K}_k^{\theta,\text{e}} & = 
\mathbf{\Theta}_k^\text{f}
\left(\mathbf{Y}_k^\text{f}\right)^\top
\left(
\mathbf{Y}_k^\text{f}
\left(\mathbf{Y}_k^\text{f}\right)^\top
+
\mathbf{E}_k^\text{o} \left(\mathbf{E}_k^\text{o}\right)^\top
\right)^{-1}\\
\mathbf{\theta}_{k}^{\text{a},(i)} & =
\mathbf{\theta}_{k}^{\text{f},(i)}+
\mathbf{K}_k^{\theta,\text{e}}
\left(
\mathbf{y}_k^{\text{o},(i)}-\mathbf{y}_{k}^{\text{f},(i)}
\right)
\end{align*}
\;
\vspace*{-9mm}
\nlset{6:}State forecast:
\vspace*{-4mm}
\begin{align*}
\mathbf{x}_{k}^{\text{f},(i)} & = \mathbf{\mathcal{M}}_{k:k-1}(\mathbf{x}_{k-1}^{\text{a},(i)},\mathbf{\theta}_{k}^{\text{a},(i)})\\
\mathbf{y}_{k}^{\text{f},(i)} & = \mathbf{\mathcal{H}}_k\left(\mathbf{x}_{k}^{\text{f},(i)}\right)
\end{align*}
\;
\vspace*{-9mm}
\nlset{7:}Compute the normalized anomalies
\vspace*{-4mm}
$$
\left[\mathbf{X}_k^\text{f}\right]_{:,i}=
\frac{\mathbf{x}_{k}^{\text{f},(i)}-\overline{\mathbf{x}_{k}^{\text{f}}}}{\sqrt{N_\text{e}-1}}
\: 
;
\:
\left[\mathbf{Y}_k^\text{f}\right]_{:,i}=
\frac{%
\mathbf{y}_{k}^{\text{f},(i)}-
\overline{\mathbf{y}_{k}^{\text{f}}}
}{\sqrt{N_\text{e}-1}}
\: 
;
\:
\left[\mathbf{E}_k^\text{o}\right]_{:,i}=
\frac{%
\mathbf{\epsilon}_k^{\text{o},(i)}-\overline{\mathbf{\epsilon}_{k}^{\text{o}}}
}{\sqrt{N_\text{e}-1}}
$$
\;
\vspace*{-9mm}
\nlset{8:}State update:
\vspace*{-4mm}
\begin{align*}
\mathbf{K}_k^{x,\text{e}} & = 
\mathbf{X}_k^\text{f}
\left(\mathbf{Y}_k^\text{f}\right)^\top
\left(
\mathbf{Y}_k^\text{f}
\left(\mathbf{Y}_k^\text{f}\right)^\top
+
\mathbf{E}_k^\text{o} \left(\mathbf{E}_k^\text{o}\right)^\top
\right)^{-1}\\
\mathbf{x}_{k}^{\text{a},(i)} & =
\mathbf{x}_{k}^{\text{f},(i)}+
\mathbf{K}_k^{x,\text{e}}
\left(
\mathbf{y}_k^{\text{o},(i)}-\mathbf{y}_{k}^{\text{f},(i)}
\right)
\end{align*}
\;
\vspace*{-9mm}
}
}
\caption{\label{alg:Stoch_Dual_EnKF_Moradkhani_Carrassi}Dual Ensemble Kalman Filter (slightly adapted from \cite{DENKF_MORADKHANI2005}). Use of anomaly matrices with $\mathbf{Y}_k^\text{f}=\mathbf{H}_k\mathbf{X}_k^\text{f}$. We have $i=1,\cdots,N_\text{e}$.}
\end{algorithm}

\subsection{Multigrid Ensemble Kalman filter algorithm}
\label{sec:MEnKF}

%
\begin{algorithm}
\setstretch{0.9}
\KwIn{For $k=1,\dots,K$: 
the forward models $\mathbf{\mathcal{M}}_{k:k-1}^\text{\tiny C}$, 
the observation models $\mathbf{\mathcal{H}}_k^\text{\tiny C}$, 
the observation error covariance matrices $\mathbf{R}_k^\text{\tiny C}$}
\KwOut{$\{\mathbf{\theta}_{k}^{\text{a},(i)}\}$ and $\{\left(\mathbf{x}_{k}^\text{\tiny C}\right)^{\text{a},(i)}\}$\,;\, $k=0,\cdots,K$}
\Begin{
\nlset{1:}Initialize $\{\mathbf{\theta}_{0}^{\text{a},(i)}\}$ and $\{\left(\mathbf{x}_{0}^\text{\tiny C}\right)^{\text{a},(i)}\}$\;
\For{$k=1,\dots,K$}{
\nlset{2:}Parameter forecast:
\vspace*{-3mm}
\begin{align*}
\mathbf{\theta}_{k}^{\text{f},(i)} & = \mathbf{\theta}_{k-1}^{\text{a},(i)}+\mathbf{\tau}_{k}^{(i)}\quad\text{with}\quad\mathbf{\tau}_{k}^{(i)}\sim \mathcal{N}(0,\mathbf{\Sigma}_k^\mathbf{\theta})\\
\left(\mathbf{x}_{k}^\text{\tiny C}\right)^{\text{f},(i)} & = \mathbf{\mathcal{M}}_{k:k-1}^\text{\tiny C}\left(\left(\mathbf{x}_{k-1}^\text{\tiny C}\right)^{\text{a},(i)},\mathbf{\theta}_{k}^{\text{f},(i)}\right)
\end{align*}
     \If {Observation available}{
     	\vspace*{-3mm}
        \begin{align*}
		\left(\mathbf{y}_{k}^\text{\tiny C}\right)^{\text{f},(i)} & = \mathbf{\mathcal{H}}_k^\text{\tiny C}\left(\left(\mathbf{x}_{k}^\text{\tiny C}\right)^{\text{f},(i)}\right)     
		\end{align*}
		\vspace*{-1mm}
		\nlset{3:}Observation ensemble:
		\vspace*{-3mm}
		\begin{align*}
		\left(\mathbf{y}_k^\text{\tiny C}\right)^{\text{o},(i)} & =\left(\mathbf{y}_k^\text{\tiny C}\right)^\text{o} + \left(\mathbf{\epsilon}_k^\text{\tiny C}\right)^{\text{o},(i)}\quad\text{with}\quad\left(\mathbf{\epsilon}_k^\text{\tiny C}\right)^{\text{o},(i)}\sim \mathcal{N}(0,\mathbf{R}_k^\text{\tiny C})\\
		\displaystyle
		\left(\mathbf{R}_k^\text{\tiny C}\right)^\text{e}
		& =
		\frac{1}{N_\text{e}-1}
		\sum_{i=1}^{N_\text{e}}
		\left(\mathbf{\epsilon}_k^\text{\tiny C}\right)^{\text{o},(i)}
		\left(
		\left(\mathbf{\epsilon}_k^\text{\tiny C}\right)^{\text{o},(i)}
		\right)^\top
		\end{align*}
		\vspace*{-3mm}
		\nlset{4:}Compute the normalized anomalies
		\vspace*{-0mm}
		$$
		\left[\mathbf{\Theta}_k^\text{f}\right]_{:,i}=
		\frac{\mathbf{\theta}_{k}^{\text{f},(i)}-\overline{\mathbf{\theta}_{k}^{\text{f}}}}{\sqrt{N_\text{e}-1}}
		\: 
		;
		\:
		\left[\mathbf{Y}_k^\text{f}\right]_{:,i}=
		\frac{%
		\left(\mathbf{y}_{k}^\text{\tiny C}\right)^{\text{f},(i)}-
		\overline{\left(\mathbf{y}_{k}^\text{\tiny C}\right)^{\text{f}}}
		}{\sqrt{N_\text{e}-1}}
		\: 
		;
		\:
		\left[\mathbf{E}_k^\text{o}\right]_{:,i}=
		\frac{%
		\left(\mathbf{\epsilon}_k^\text{\tiny C}\right)^{\text{o},(i)}-\overline{\left(\mathbf{\epsilon}_{k}^\text{\tiny C}\right)^{\text{o}}}
		}{\sqrt{N_\text{e}-1}}
		$$
		\vspace*{-3mm}
		\nlset{5:}Parameter update:
		\vspace*{-1mm}
		\begin{align*}
		\left(\mathbf{K}_k^\text{\tiny C}\right)^{\theta,\text{e}} & = 
		\mathbf{\Theta}_k^\text{f}
		\left(\mathbf{Y}_k^\text{f}\right)^\top
		\left(
		\mathbf{Y}_k^\text{f}
		\left(\mathbf{Y}_k^\text{f}\right)^\top
		+
		\mathbf{E}_k^\text{o} \left(\mathbf{E}_k^\text{o}\right)^\top
		\right)^{-1}\\
		\mathbf{\theta}_{k}^{\text{a},(i)} & =
		\mathbf{\theta}_{k}^{\text{f},(i)}+
		\left(\mathbf{K}_k^\text{\tiny C}\right)^{\theta,\text{e}}
		\left(
		\left(\mathbf{y}_k^\text{\tiny C}\right)^{\text{o},(i)}-\left(\mathbf{y}_{k}^\text{\tiny C}\right)^{\text{f},(i)}
		\right)
		\end{align*}
		\nlset{6:}State forecast:
		\vspace*{-5mm}
		\begin{align*}
		\left(\mathbf{x}_{k}^\text{\tiny C}\right)^{\text{f},(i)} & = \mathbf{\mathcal{M}}_{k:k-1}^\text{\tiny C}\left(\left(\mathbf{x}_{k-1}^\text{\tiny C}\right)^{\text{a},(i)},\mathbf{\theta}_{k}^{\text{a},(i)}\right)\\
		\left(\mathbf{y}_{k}^\text{\tiny C}\right)^{\text{f},(i)} & = \mathbf{\mathcal{H}}_k^\text{\tiny C}\left(\left(\mathbf{x}_{k}^\text{\tiny C}\right)^{\text{f},(i)}\right)
		\end{align*}
		\vspace*{-3mm}
		\nlset{7:}Compute the normalized anomalies
		\vspace*{-1mm}
		$$
		\left[\mathbf{X}_k^\text{f}\right]_{:,i}=
		\frac{\left(\mathbf{x}_{k}^\text{\tiny C}\right)^{\text{f},(i)}-\overline{\left(\mathbf{x}_{k}^\text{\tiny C}\right)^{\text{f}}}}{\sqrt{N_\text{e}-1}}
		\: 
		;
		\:
		\left[\mathbf{Y}_k^\text{f}\right]_{:,i}=
		\frac{%
		\left(\mathbf{y}_{k}^\text{\tiny C}\right)^{\text{f},(i)}-
		\overline{\left(\mathbf{y}_{k}^\text{\tiny C}\right)^{\text{f}}}
		}{\sqrt{N_\text{e}-1}}
		\: 
		;
		\:
		\left[\mathbf{E}_k^\text{o}\right]_{:,i}=
		\frac{%
		\left(\mathbf{\epsilon}_k^\text{\tiny C}\right)^{\text{o},(i)}-\overline{\left(\mathbf{\epsilon}_{k}^\text{\tiny C}\right)^{\text{o}}}
		}{\sqrt{N_\text{e}-1}}
		$$
		\vspace*{-5mm}
		\nlset{8:}State update:
		\vspace*{-0mm}
		\begin{align*}
		\left(\mathbf{K}_k^\text{\tiny C}\right)^{x,\text{e}} & = 
		\mathbf{X}_k^\text{f}
		\left(\mathbf{Y}_k^\text{f}\right)^\top
		\left(
		\mathbf{Y}_k^\text{f}
		\left(\mathbf{Y}_k^\text{f}\right)^\top
		+
		\mathbf{E}_k^\text{o} \left(\mathbf{E}_k^\text{o}\right)^\top
		\right)^{-1}\\
		\left(\mathbf{x}_{k}^\text{\tiny C}\right)^{\text{a},(i)} & =
		\left(\mathbf{x}_{k}^\text{\tiny C}\right)^{\text{f},(i)}+
		\left(\mathbf{K}_k^\text{\tiny C}\right)^{x,\text{e}}
		\left(
		\left(\mathbf{y}_k^\text{\tiny C}\right)^{\text{o},(i)}-\left(\mathbf{y}_{k}^\text{\tiny C}\right)^{\text{f},(i)}
		\right)
		\end{align*}
     } 
} 
} 
   \caption{Dual Ensemble Kalman filter of Algo.~\ref{alg:Stoch_Dual_EnKF_Moradkhani_Carrassi} applied on the coarse mesh. We have $i=1,\cdots,N_\text{e}$.}
   \label{alg:Stoch_Dual_EnKF_Moradkhani_Carrassi_Coarse_Grid}
\end{algorithm}

\begin{algorithm}
\DontPrintSemicolon
\Begin{
\nlset{1:}Initialize 
$\left\{
\left(\mathbf{x}_{0}^\text{\tiny F}\right)^\text{a},
\overline{\mathbf{\theta}_{0}^\text{a}},
\mathbf{\theta}_{0}^{\text{a},(i)},
\left(\mathbf{x}_{0}^\text{\tiny C}\right)^{\text{a},(i)}
\right\}$\;
\For{$k=1,\dots,K$}{
\nlset{2:}Fine grid forecast:
	$$
	\left(\mathbf{x}_k^\text{\tiny F}\right)^\text{f}=
	\mathbf{\mathcal{M}}^\text{\tiny F}_{k:k-1}\left(
	\left(\mathbf{x}_{k-1}^\text{\tiny F}\right)^\text{a},
	\overline{\mathbf{\theta}_{k}^\text{a}}\right)
	$$
\nlset{3:}Dual EnKF on coarse mesh: Apply Algo.~\ref{alg:Stoch_Dual_EnKF_Moradkhani_Carrassi_Coarse_Grid}\\
      \If {Observation available}{
\nlset{4:}Projection on the coarse grid 
$$
\left(\mathbf{x}^\text{\tiny C}_k\right)^{*}=
\Pi_\text{\tiny C}\left(\left(\mathbf{x}_k^\text{\tiny F}\right)^\text{f}\right)
$$
\nlset{5:}Fine grid state correction using the ensemble statistics:\\
		\begin{align*}
\left(\mathbf{x}^\text{\tiny C}_k\right)^{'} & =
\left(\mathbf{x}^\text{\tiny C}_k\right)^{*}+
\left(\mathbf{K}_k^\text{\tiny C}\right)^{x,\text{e}}
\left[
\left(\mathbf{y}_k^\text{\tiny C}\right)^\text{o}-
\mathbf{\mathcal{H}}_k^\text{\tiny C}
\left(\left(\mathbf{x}_k^\text{\tiny C}\right)^{*}\right)
\right]\\
\left(\mathbf{x}^\text{\tiny F}_k\right)^\text{a} & =
\left(\mathbf{x}^\text{\tiny F}_k\right)^\text{f}+
\Pi_\text{\tiny F}
\left(\left(
\mathbf{x}^\text{\tiny C}_k\right)^{'}-
\left(\mathbf{x}^\text{\tiny C}_k\right)^{*}\right) 	
		\end{align*}
\nlset{6:} Matrix-Splitting iterative procedure on the final solution starting from $\left(\mathbf{x}^\text{\tiny F}_k\right)^\text{a}$. 
			} 
} 
} 
   \caption{Multigrid EnKF algorithm. We have $i=1,\cdots,N_\text{e}$.}
   \label{alg:algo_KF_corrected_corrected}
\end{algorithm}


The algorithm \ref{alg:algo_KF_corrected_corrected} represents a simplified, ready-to-use application of the conceptual methodology presented in Sec.~\ref{sec:multigrid-EnKF}. This algorithm was tailored for the relatively simple physical models used in this work. While it may not be suited for complex three-dimensional applications, it proved an optimum trade-off in accuracy and computational resources for the present analysis. 

First of all, when observation is not available, the two main forecast operations (fine grid forecast and ensemble coarse forecast) are performed using explicit time advancement schemes. This choice allows to reduce the computational costs. However, when observation is available, the following strategies are employed: 
\begin{enumerate}
\item The two forecast operations (main simulation and ensemble members) are performed using an implicit matrix-splitting iterative procedure, using a single iteration. As stated in Sec.~\ref{sec:multigrid-EnKF}, the state transition model for each ensemble member is determined independently, but using the same structure and discretization schemes of the main simulation. 
\item The number of iterative solutions for the main simulation on the coarse-grid level is equal to zero. That is, the solution from the first forecast is projected on the coarse grid, and the difference between the KF state estimation and this forecast is re-projected over the fine grid.
\item In the final iteration on the fine grid, an implicit matrix-splitting iterative procedure is employed, using a single iteration and a relaxation coefficient $\alpha=0.5$. This choice, which provides the best compromise between accuracy and regularization, has been identified after extensive tests for the configurations investigated.

\end{enumerate}

\bibliography{Bibliography_HDR}


\end{document}